\documentclass[5p]{elsarticle}
\usepackage{graphicx}
\usepackage{multicol}
\usepackage{amssymb}
\usepackage{amsthm}
\usepackage{amsmath}
\usepackage{float}
\usepackage{multirow}
\usepackage{algorithmic}
\usepackage{algorithm}
\usepackage{lineno}
\usepackage{url}
\usepackage{subfig}
\usepackage{diagbox}
\usepackage{amsmath,amsfonts,amsthm,bm} 
\usepackage[usenames,dvipsnames]{xcolor}
\usepackage{setspace}
\usepackage{lipsum}

\usepackage[final]{listingsutf8}  

\usepackage[final]{listingsutf8}  

\journal{arXiv.org}

\begin{document}

\begin{frontmatter}


\title{Bayesian inference under small sample size -- A noninformative prior approach}


\author[buaa]{Jingjing He}
\address[buaa]{School of Reliability and Systems Engineering, Beihang University, 37 Xueyuan Rd., Beijing 100191, China}

\author[gscaep]{Xuefei Guan\corref{corauthor}}
\address[gscaep]{Graduate School of China Academy of Engineering Physics, 10 Xibeiwang E. Rd., Beijing 100193, China}

\cortext[corauthor]{Corresponding author, Email: xfguan@gscaep.ac.cn}

\begin{abstract}
A Bayesian inference method for problems with small samples and sparse data is presented in this paper. A general type of prior ($\propto 1/\sigma^{q}$) is proposed to formulate the Bayesian posterior for inference problems under small sample size. It is shown that this type of prior can represents a broad range of priors such as classical noninformative priors and asymptotically locally invariant priors. It is further shown in this study that such priors can be derived as the limiting states of Normal-Inverse-Gamma conjugate priors, allowing for analytical evaluations of Bayesian posteriors and predictors. The performance of different noninformative priors under small sample size is compared using the global likelihood. The method of Laplace approximation is employed to evaluate the global likelihood. A numerical linear regression problem and a realistic fatigue reliability problem are used to demonstrate the method and identify the optimal noninformative prior. Results indicate the predictor using Jeffreys' prior outperforms others. The advantage of the noninformative Bayesian estimator over the regular least square estimator  under small sample size is shown.
\end{abstract}


\end{frontmatter}

%

\section{Introduction}
\label{sec:intro}

Sample sizes in a vast array of engineering fields are frequently quite small due to time, economic, and physical constraints. For example, life testing data of high-reliability mechanical components, large and complex engineering systems, and so on. The effect of sample size on the interpretation of classical significance tests has been emphasized in several studies \cite{royall1986effect,raudys1991small,nelson1993predictable}. In addition, the predictive model built upon a small number of samples may highly depend on the chosen method for parameter estimation \cite{fan1999effects,mcneish2016using}. To enable meaningful model prediction and results interpretation, the probabilistic approach is usually preferred over the deterministic approach \cite{winkler1971probabilistic,melchers2018structural,he2020lifetime}. The Bayes rule provides a consistent and rational mathematical device to incorporate relevant information and prior knowledge for probabilistic inference.

To motivate the discussion, consider an observable random variable $X$ with a conditional probability density function (PDF, or simply density) of $p(x|\phi)$, where $\phi\in\Phi$ is also a random variable. The inverse problem is to make inferences about $\phi$ given an observed value $x$ of $X$. The Bayesian approach to the solution is to use some density $p(\phi)$ over $\Phi$ to represent the prior information of $\phi$. In this way, the prior knowledge of $\phi$ can be encoded through the Bayes rule to obtain the posterior PDF of $\phi$ given $x$, 
\begin{equation}
p(\phi|x) \propto p(\phi)p(x|\phi).
\label{eq:inverse}
\end{equation}
The forward inference, e.g., the PDF of a certain variable or the probability of an event involving $\phi$, can be made on the basis of the posterior PDF. Using the method of Markov chain Monte Carlo (MCMC), samples can directly be drawn from the posterior distribution without knowing the normalizing constant \allowbreak{$p(x)=\int p(\phi)p(x|\phi)\mathrm{d}\phi$} in Eq. (\ref{eq:inverse}). The Bayesian method has been successfully demonstrated in all important disciplines \cite{Gregory2005,hamdia2016fracture,yang2016probabilistic,guan2019life,wang2018model,chen2018equivalent,he2020lamb,huo2020bayesian}. 

The choice of $p(\phi)$ can have a great influence on the inference result. The proper choice of priors has been extensively discussed in probability and statistics communities, and it can never be overlooked as it is one of the fundamental pieces in Bayesian inference \cite{kass1996selection,gelman2017beyond}. For one thing, the formal rule of constructing a prior regardless of the data and likelihood is sought in the field of physics \cite{jaynes1968prior, du2020maximum}. \citet{Jaynes2003} argued that a problem of inference is ill-posed until three essential things are recognized: the prior probabilities represent one's prior information, and are to be determined, not by introspection but by logical analysis of that information; one must specify the prior information to be used just as fully as one specifies the data in formulating a problem; and the goal is that inferences are to be completely `objective' in the sense that two persons with the same prior information must assign the same prior probabilities. For another, the choice of a prior may highly depend on data and likelihood in practice. \citet{gelman2017prior} argued that a prior can in general only be interpreted in the context of the likelihood with which it will be paired. 

To ensure a consistent and objective inference, rules for constructing priors with minimal subjective constraints are sought. Early work on construction of such priors is based on the `ignorance' over the parameter space using invariance techniques \cite{jeffreys1946invariant, jaynes1968prior,hartigan1964invariant}. The fundamental reasoning is that the priors should carry the same amount of information such that a change of scale and/or a shift of location do not affect the inference results on those parameters. The ignorant prior can systematically be derived using the concept of transformation group. Using different transformation groups different priors can be obtained. For simplicity, such priors are loosely referred to as noninformative priors. One of the most notable priors for a scale parameter $\sigma$ of a distribution is Jeffreys' prior, i.e., $p(\sigma)\propto 1/\sigma$. Jeffreys' prior can be obtained using the tool of transformation group under the condition that a change of scale does not change that state of knowledge. A further extension of the noninformative priors are ``reference priors'' \cite{bernardo1979reference,berger1992ordered}. The theoretical framework has been discussed in many studies including, but not limited to, Refs. \cite{wigner2012group, hartigan1964invariant, jeffreys1998theory}. Apart from noninformative approach to derive the priors, there are several less objective approaches to construct the priors. \citet{gelman2006prior} proposed the weakly informative priors based on the idea of conditional conjugacy for hierarchical Bayesian models. The conditionally conjugate priors provide some computational convenience as a Gibbs sampler can be used to draw samples from a posterior distribution, and some parameters for Inverse--Gamma distributions in Bayesian applications are suggested. \citet{simpson2017penalising} proposed a method to build priors. The basic idea is to penalize the complexity induced by deviating from a simpler base model using the Kullback--Leiber divergence as a metric \cite{kullback1951information}. 

For critical problems with small sample sizes, the choice of the prior in Bayesian inference can have a great impact on the inference results, rendering an unreliable decision-making. Despite a great deal of research, including the aforementioned, has been conducted, the choice of an optimal prior is still nontrivial to make from the practical point of view, and a systematical method to cope with such cases is rarely seen. Moreover a quantitative measure to evaluate the performance of a prior is equally important to identify the optimal prior. This study develops a noninformative Bayesian approach to probabilistic inference under small sample size. A general $1/\sigma^{q}$-type of noninformative prior for the location-scale family of distributions is proposed. This type of prior is further shown to be the limiting state of the commonly-used Normal--Inverse--Gamma conjugate prior, and therefore allowing for analytical evaluation of the posterior of the parameter. Given a linear model with a Gaussian error variable, the analytical form of the posterior of the model prediction can also be obtained. 
The remainder of the paper is organized as follows. First, the Bayesian model with noninformative priors are developed, in particular, a general $1/\sigma^{q}$-type of noninformative prior for the location-scale family of distributions is proposed for small sample problems. The $1/\sigma^{q}$-type of noninformative priors are further shown as the limiting states of the Normal--Inverse--Gamma (NIG) conjugate priors. The close--form expressions of the Bayesian posterior and predictors using the proposed noninformative priors are obtained. Next, a generic performance measure considering both the fitting performance and predictive performance is proposed using the concept of Bayes factors. Different priors are treated as models in a Bayesian hypothesis testing context for comparisons. Following that, the overall method is demonstrated using a simple linear regression problem and a fatigue reliability problem under small sample. The comparisons of the developed Bayesian approach with the regular least square method are made. Finally conclusions are drawn based on the current study.

\section{Bayesian linearized models with noninformative priors}

To motivate the discussion, consider a general linear or linearized model
\begin{equation}
y_{i} = \mathbf{x}_{i}\bm{\theta}+\epsilon_{i},
\label{eq:linear}
\end{equation}
where $\bm{\theta}$ is a $k$-dimensional column vector, and $\epsilon_{i}$ are independent and identical distributed random error variables. The distribution of $\epsilon_{i}$ determines the likelihood function or vice versa. Without loss of generality, the distribution belongs to a location-scale family of distributions. Furthermore it is enough to use a single scale parameter to characterize the distribution of $\epsilon$ since any constant non-zero mean, no matter known or unknown, can be grouped into $\bm{\theta}$. Denote the scale parameter of $\epsilon$ as $\sigma^2$. 
A Bayesian model incorporates both the prior information and the observation through Bayes rule. Using the matrix form $\mathbf{y}=\mathbf{x}\bm\theta+\bm\epsilon$, the Bayesian posterior of $(\bm{\theta},\sigma^2)$ writes
\begin{equation}
p({\bm\theta},\sigma^2|{\bf y}) \propto p({\bm\theta},\sigma^2) p({\bf y}|{\bm\theta},\sigma^2).
\label{eq:bayesian}
\end{equation}
The common Gaussian error variable, i.e., $\epsilon_{i}\sim N(0,\sigma^{2})$, corresponds to the following likelihood function, for $n$ observations $\mathbf{y}=(y_{1},y_{2},...,y_{n})^{T}$ and $n$ input vector $\mathbf{x}=(\mathbf{x}_{1},\mathbf{x}_{2},...,\mathbf{x}_{n})^{T}$ where $\mathbf{x}_{i}$, $i=1,...,n$, is a row vector of size $k$.
\begin{equation}
\begin{array}{rl}
p(\mathbf{y}|\bm{\theta},\sigma^{2}) & =\mathrm{N}(\mathbf{x}\bm{\theta},\sigma^{2}\mathrm{I}) \\
& \propto \sigma^{-n}\exp\left[-\displaystyle\frac{1}{2\sigma^2}\sum_{i=1}^{n}\left(y_{i}-\mathbf{x}_{i}\bm{\theta} \right)^2 \right].
\end{array}
\label{eq:likelihood}
\end{equation}
It should be noted that the error variable does not necessarily follow a Gaussian PDF, and other types of error distributions can be used. For example, the extreme-value PDF for the error corresponds to a Weibull likelihood for the log-transformed model prediction. 

\subsection{A general form of noninformative priors -- $\propto 1/\sigma^{q}$ }

A general $\propto 1/\sigma^{q}$ ($q>=0$) form of priors are considered here. The classical Jeffreys prior and the related asymptotically locally invariant prior are introduced first for the purpose of completeness. 

Consider the PDF of a random variable $x$ characterized by a parameter vector $\bm{\phi}$, Jeffreys' noninformative prior distribution of $\bm{\phi}$ is proportional to the square root of the determinant of the Fisher information matrix, e.g., 

\begin{equation}
p(\bm{\phi}) \propto \sqrt{\mathrm{det}\mathbf{I}(\bm{\phi)}},
\label{eq:Jefferyprior}
\end{equation}
where $\mathrm{det}(\cdot)$ is the determinant operator, and $\mathbf{I}(\cdot)$ is the Fisher information matrix. The key feature of it is invariance under monotone transformation of $\bm{\phi}$. This feature is achieved by using the change of variables theorem. Denote the reparameterized variable or vector as $\bm{\psi}$, it can be shown that
\begin{equation}
p(\bm{\psi}) = \displaystyle p(\bm{\phi})\left|\mathrm{det}\frac{\partial \phi_{i}}{\partial \psi_{j}} \right| = \sqrt{\mathrm{det}\mathbf{I}(\bf{\psi})},
\label{eq:reparvec}
\end{equation}
For a Gaussian likelihood with unknown parameters $\mu$ and $\sigma^2$,
\begin{equation}
p(x|\mu,\sigma^2)=\frac{1}{\sqrt{2\pi\sigma^2}}\exp\left[-\frac{(x-\mu)^2}{2\sigma^2} \right].
\label{eq:gauss}
\end{equation}
The Jeffreys prior for the joint parameter $(\mu,\sigma^2)$ is
\begin{equation}
p(\mu,\sigma^2) \propto \sqrt{\mathrm{det}\mathbf{I}(\mu,\sigma^2)},
\label{eq:}
\end{equation}
where $\mathbf{I}(\mu,\sigma^2)$ is
\begin{equation}
\mathbf{I} = \mathbb{E}
\begin{bmatrix}
\dfrac{\partial\ln p(x|\cdot)}{\partial\mu}\dfrac{\partial\ln p(x|\cdot)}{\partial \mu} && \dfrac{\partial\ln p(x|\cdot)}{\partial\mu}\dfrac{\partial\ln p(x|\cdot)}{\partial (\sigma^2)} \\
\dfrac{\partial\ln p(x|\cdot)}{\partial\mu}\dfrac{\partial\ln p(x|\cdot)}{\partial (\sigma^2)} &&\dfrac{\partial\ln p(x|\cdot)}{\partial (\sigma^2) }\dfrac{\partial\ln p(x|\cdot)}{\partial (\sigma^2)}
\end{bmatrix}
\label{eq:hmat}
\end{equation}
and $\mathbb{E}(\cdot)$ is the expectation operator.
Using algebraic deduction and integration, $\mathbf{I}(\mu,\sigma^2)$ is simplified to
\begin{equation}
\mathbf{I} =\left[ 
\begin{matrix}
\dfrac{1}{\sigma^2} && 0 \\
0 && \dfrac{2}{\sigma^{4}}
\end{matrix}
\right].
\label{eq:hmaxs}
\end{equation}
As a result the Jeffreys prior for $(\mu,\sigma^2)$ is  
\begin{equation}
p(\mu,\sigma^2) \propto 1/\sigma^3.
\label{eq:joint1}
\end{equation}
It is noted that the distribution of interest is $(\mu,\sigma^2)$, not $(\mu,\sigma)$; therefore, the derivative is taken with respective to $\sigma^2$ as a whole instead of $\sigma$. In other word, the parameter space is on $\sigma^2$, not $\sigma$. For example, the Jeffreys prior for $\sigma^2$, with a fixed value of $\mu$, is 
\begin{equation}
p(\sigma^2) \propto \frac{1}{\sigma^2}. 
\label{eq:joint2}
\end{equation}
The Jeffreys priors for $(\mu,\sigma)$ and $\sigma$ with a fixed $\mu$ are $\propto 1/\sigma^2$ and $\propto 1/\sigma$, respectively. It is shown in Appendix \ref{app:fishkl} that the Fisher information matrix is also the Hessian matrix of the Kullback--Leibler (KL) distance of a deviated distribution with respect to the true distribution evaluated at the true parameters. For exponential families of distributions, the KL distance has analytical forms, allowing for the evaluation of the Fisher information matrix without involving integrals in Eq. (\ref{eq:hmat}).

The asymptotically locally invariant (ALI) prior is another type of priors that satisfy the invariance under certain transformations. An ALI prior can be uniquely determined using the following equation according to Ref. \cite{hartigan1964invariant},
\begin{equation}
\frac{\partial\ln p(\phi)}{\partial\phi} = -\mathbb{E}\left[f_{1}f_{2} \right]/\mathbb{E}\left[ f_{2}\right],
\label{eq:par}
\end{equation}
where $f_{1} = \partial\ln p(x|\cdot) / \partial\phi$ and $f_{2} = \partial^2\ln p(x|\cdot) / \partial\phi^2$. For a Gaussian distribution with a fixed mean and a random $\sigma$, the two terms are $\mathbb{E}\left[f_{1}f_{2}\right]=-6/\sigma^{3}$, and $\mathbb{E}\left[f_{2}\right]=-2/\sigma^{2}$. Solving 
\begin{equation}
\frac{\partial \ln p(\sigma)}{\partial\sigma}=-\frac{3}{\sigma}
\label{eq:df}
\end{equation}
to obtain the ALI prior for $\sigma$
\begin{equation}
p(\sigma)\propto \frac{1}{\sigma^{3}}.
\label{eq:hsigma}
\end{equation}
Similarly, the ALI prior for $\sigma^{2}$ is
\begin{equation}
p(\sigma^2)\propto \frac{1}{\sigma^{4}}.
\label{eq:alisigma2}
\end{equation}
When both $\mu$ and $\sigma$ are random, the joint ALI prior for $(\mu,\sigma)$ is
\begin{equation}
p(\mu,\sigma)\propto\frac{1}{\sigma^{5}}.
\label{eq:alimusigma}
\end{equation}
It is noticed that the Jeffreys, ALI, and uniform priors are reproduced from $\propto 1/\sigma^{q}$ as $q$ takes different integer values. In the following the derivation of a $1/\sigma^{q}$ prior from NIG conjugates is shown.

\subsection{Derivation of the $1/\sigma^{q}$ priors as the limiting states of NIG conjugates}

In Bayesian models if the posterior distributions $p(\phi|x)$ are in the same family of distributions as the prior $p(\phi)$, the prior and posterior are called conjugate distributions. The prior is referred to as a conjugate prior for the given likelihood function. It can be considered as the prior can be reconditioned by encoding the evidence through the likelihood; therefore, the evidence or data merely change the distribution parameters of the prior and yield another distribution of the same type but with a different set of parameters. 

For a location--scale parameter vector $({\bm\theta},\sigma^2)$ used in linear or linearized Bayesian models with a Gaussian likelihood, the corresponding conjugate prior is the NIG distribution. The posterior PDF for the parameter and prediction are given in closed forms are given in Appendix \ref{apdx:conjugate}. 

Noninformative priors, including Jeffreys, ALI, and reference priors, for $({\bm\theta},\sigma^2)$ are mostly in the form of $1/\sigma^{q}$, $q\in{1,2,...}$. The uniform prior can be seen as a special case of $1/\sigma^{q}$ as $q=0$. It is shown as follows that $1/\sigma^{q}$ type of noninformative priors are obtained as certain limiting states of NIG conjugates of $(\bm{\theta},\sigma^2)$. For example, the NIG distribution with parameters $(\alpha,\beta,{\bm\Sigma},{\bm\mu})$ given by Eq. (\ref{eq:nig}) can reduce to the Jeffreys prior 
\begin{equation}
p(\bm{\theta},\sigma^2) \to \frac{1}{\sigma^2}
\label{eq:NIGlimit}
\end{equation}
as
\begin{equation}
\left\{
\begin{array}{rl}
\alpha& \to -k/2 \\
\beta& \to 0^{+}\\
\bm{\Sigma}^{-1}&\to \mathbf{0} \\
\lvert{\bm\mu}\rvert & < \infty
\end{array}
\right. .
\label{eq:nigjef}
\end{equation}
Furthermore, $1/\sigma^{q}$ type of priors can all be derived as reduced NIG distributions. By assigning the initial values for $\alpha$, $\beta$, and $\bm\Sigma$ of the NIG distribution, different $q$ values are obtained. Table \ref{tab:prior_class} presents priors with different $q$ values as reduced NIG distributions and the corresponding $\alpha$, $\beta$, and $\bm\Sigma$ of the NIG distributions, and $\alpha^{*}$, $\beta^{*}$, and $\bm{\Sigma}^{*}$ of the resulting NIG posterior distributions. The posterior distribution of $\sigma^2$ is an inverse gamma distribution $\mathrm{IG}(\alpha^{*},\beta^{*})$ and the PDF is given by
\begin{equation}
p(\sigma^2|\mathbf{y}) = \frac{\beta^{*\alpha^{*}}}{\Gamma(\alpha^{*})}\left(\frac{1}{\sigma^2} \right)^{\alpha^*+1}\exp\left(-\frac{\beta^{*}}{\sigma^{2}} \right).
\label{eq:postsigma2}
\end{equation}
The posterior distribution of $\bm\theta$ is obtained by integrating out $\sigma^2$ from the joint posterior distribution of Eq. (\ref{eq:postpdf}) as,
\begin{equation}
\resizebox{1.0\columnwidth}{!}{$
\begin{aligned}
p(\bm{\theta}|\mathbf{y}) &= \int p(\bm{\theta},\sigma^2|\mathbf{y})\mathrm{d}\sigma^2 \\
&\propto \int \left(\frac{1}{\sigma^2}\right)^{\alpha^* +1 }\exp\left\{-\frac{1}{\sigma^2}\left[\beta^* + \frac{1}{2}\left(\bm{\theta}-\bm{\mu}^{*} \right)^{T}\bm{\Sigma}^{*-1}\left(\bm{\theta}-\bm{\mu}^{*} \right) \right] \right\}\mathrm{d}\sigma^2 \\
&\propto\left[1+\frac{1}{2\beta^{*}} \left(\bm{\theta}-\bm{\mu}^{*} \right)^{T}\bm{\Sigma}^{*-1}\left(\bm{\theta}-\bm{\mu}^{*} \right)\right]^{-(2\alpha^{*}+k)/2},
\end{aligned}
$}
\label{eq:posttheta}
\end{equation}
which is a $(2\alpha^{*})$ degrees-of-freedom multivariate {\it t}-- distribution with a location vector of $\bm{\mu}^{*}$ and a shape matrix of $\left(\frac{\beta^{*}}{\alpha^{*}} \right)\bm{\Sigma^{*}}$. 
In particular, when the NIG prior for $(\bm{\theta},\sigma^2)$ is reduced to the Jeffreys prior $1/\sigma^2$ by Eq. (\ref{eq:nigjef}), the resulting posteriors of $\sigma^2$ and $\bm\theta$ are
\begin{equation}
p(\sigma^{2}|\mathbf{y}) = \mathrm{IG}\left(\frac{n-k}{2},\;\frac{\mathrm{SSE}}{2}\right),
\label{eq:postsigma2theory}
\end{equation}
and
\begin{equation}
p(\bm{\theta}|\mathbf{y}) = \mathrm{MVT}_{n-k}\left( (\mathbf{x}^{T}\mathbf{x})^{-1}(\mathbf{x}^{T}\mathbf{y}),\; \frac{\mathrm{SSE}}{n-k} (\mathbf{x}^{T}\mathbf{x})^{-1} \right),
\label{eq:postthetatheory}
\end{equation}
respectively. The term $\mathrm{SSE}$ in Eq. (\ref{eq:postsigma2theory}) is the sum of squared errors, given by
\begin{equation}
\mathrm{SSE}=\sum_{i}^{n}\left( y_{i}-\mathbf{x}_{i}\bm{\theta}\right)^2.
\label{eq:SSE}
\end{equation}
The prediction posterior in this case is 
\begin{equation}
\resizebox{1.0\columnwidth}{!}{$
p(\tilde{\mathbf{y}}|{\mathbf{y}}) = \mathrm{MVT}_{n-k}\left( \tilde{\mathbf{x}}(\mathbf{x}^{T}\mathbf{x})^{-1}(\mathbf{x}^{T}\mathbf{y}),\;\frac{\mathrm{SSE}}{n-k}\left(\mathbf{I} + \tilde{\mathbf{x}}(\mathbf{x}^{T}\mathbf{x})^{-1}\tilde{\mathbf{x}} \right)
\right).
\label{eq:posty}
$}
\end{equation}
\begin{table*}[ht]%
\centering
\caption{$1/\sigma^{q}$ type of priors as limiting state cases of NIG distribution and its NIG posteriors.}
\begin{tabular}{lll}
Prior $(\theta,\sigma^2)$ & NIG$(\alpha,\beta,\bm{\theta},\sigma^{2})\to$ Prior & Posterior $\to$ NIG$(\alpha^{*},\beta^{*},\bm{\theta},\sigma^{2})$ \\
\hline
flat & $\begin{array}{rl} \alpha &= -k/2-1 \\ \beta & \to 0 \\ \bm{\Sigma}^{-1} & \to \bm{0} \end{array}$ &  $\begin{array}{rl} \alpha^{*} &= \alpha+n/2 = (n-k-2)/2 \\ \beta^{*} &= \beta+\mathrm{SSE}/2 = \mathrm{SSE}/2\\ \bm{\Sigma}^{*} &= \left(\bm{\Sigma}^{-1}+\mathbf{x}^{T}\mathbf{x} \right)^{-1} = \left(\mathbf{x}^{T}\mathbf{x} \right)^{-1} \end{array}$ \\ 

$\frac{1}{\sigma}$ & $\begin{array}{rl} \alpha &= -k/2-1/2 \\ \beta & \to 0 \\ \bm{\Sigma}^{-1} & \to \bm{0} \end{array}$ &  $\begin{array}{rl} \alpha^{*} &= \alpha+n/2 = (n-k-1)/2 \\ \beta^{*} &= \beta+\mathrm{SSE}/2 = \mathrm{SSE}/2 \\ \bm{\Sigma}^{*} &= \left(\bm{\Sigma}^{-1}+\mathbf{x}^{T}\mathbf{x} \right)^{-1} = \left(\mathbf{x}^{T}\mathbf{x} \right)^{-1} \end{array}$ \\ 

$\frac{1}{\sigma^{2}}$ & $\begin{array}{rl} \alpha &= -k/2 \\ \beta & \to 0 \\ \bm{\Sigma}^{-1} & \to \bm{0} \end{array}$ &  $\begin{array}{rl} \alpha^{*} &= \alpha+n/2 = (n-k)/2 \\ \beta^{*} &= \beta+\mathrm{SSE}/2 = \mathrm{SSE}/2\\ \bm{\Sigma}^{*} &= \left(\bm{\Sigma}^{-1}+\mathbf{x}^{T}\mathbf{x} \right)^{-1} = \left(\mathbf{x}^{T}\mathbf{x} \right)^{-1} \end{array}$ \\ 

$\frac{1}{\sigma^{3}}$ & $\begin{array}{rl} \alpha &= -k/2+1/2 \\ \beta & \to 0 \\ \bm{\Sigma}^{-1} & \to \bm{0} \end{array}$ &  $\begin{array}{rl} \alpha^{*} &= \alpha+n/2 = (n-k+1)/2 \\ \beta^{*} &= \beta+\mathrm{SSE}/2 = \mathrm{SSE}/2\\ \bm{\Sigma}^{*} &= \left(\bm{\Sigma}^{-1}+\mathbf{x}^{T}\mathbf{x} \right)^{-1} = \left(\mathbf{x}^{T}\mathbf{x} \right)^{-1}\end{array}$ \\ 

$\frac{1}{\sigma^{4}}$ & $\begin{array}{rl} \alpha &= -k/2+1 \\ \beta & \to 0 \\ \bm{\Sigma}^{-1} & \to \bm{0} \end{array}$ &  $\begin{array}{rl} \alpha^{*} &= \alpha+n/2 = (n-k+2)/2 \\ \beta^{*} &= \beta+\mathrm{SSE}/2 = \mathrm{SSE}/2\\ \bm{\Sigma}^{*} &= \left(\bm{\Sigma}^{-1}+\mathbf{x}^{T}\mathbf{x} \right)^{-1} = \left(\mathbf{x}^{T}\mathbf{x} \right)^{-1} \end{array}$ \\ 

$\frac{1}{\sigma^{5}}$ & $\begin{array}{rl} \alpha &= -k/2+3/2 \\ \beta & \to 0 \\ \bm{\Sigma}^{-1} & \to \bm{0} \end{array}$ &  $\begin{array}{rl} \alpha^{*} &= \alpha+n/2 = (n-k+3)/2\\ \beta^{*} &= \beta+\mathrm{SSE}/2 = \mathrm{SSE}/2 \\ \bm{\Sigma}^{*} &= \left(\bm{\Sigma}^{-1}+\mathbf{x}^{T}\mathbf{x} \right)^{-1} = \left(\mathbf{x}^{T}\mathbf{x} \right)^{-1}\end{array}$ \\ 
\hline
\end{tabular}
\label{tab:prior_class}
\end{table*} 

The advantage of treating $1/\sigma^{q}$ type of noninformative priors as reduced NIG conjugates is that the Bayesian posteriors of the model prediction and parameters all have analytical forms, allowing for efficient evaluations without resorting to the MCMC techniques as done in regular Bayesian analysis.

\section{Assessment of priors}
To evaluate the performance of priors, the Bayesian model assessment method using efficient asymptotic approximations is proposed. The idea is to recast the assessment to a model comparison problem. The participating models here are the posteriors obtained with different priors. The comparison can then be made using Bayes factors in a Bayesian hypothesis testing context. 

\subsection{Fitting performance}
The Bayes factor, on the basis of observed data $\bf{y}$, evaluating the plausibility of two different models, $M_1$ and $M_2$, can be expressed as
\begin{equation}
B_{12} = \frac{P({\bf y}|M_1)}{P({\bf y}|M_2)} = \frac
{\int p({\bf y}|{\bm\theta},\sigma^2,M_{1})p({\bm\theta},\sigma^{2}|M_{1})\mathrm{d}{\bm\theta} \mathrm{d}\sigma^2}
{\int p({\bf y}|{\bm\theta},\sigma^2,M_{2})p({\bm\theta},\sigma^{2}|M_{2})\mathrm{d}{\bm\theta} \mathrm{d}\sigma^2} 
\label{eq:B12}
\end{equation}
The comparison of two models can be made based on ratio of the posterior probabilities of the models
\begin{equation}
\frac{P(M_{1} | {\bf y})}{P(M_{2}| {\bf y})} = \frac{P({\bf y}|M_1) P(M_{1})}{P({\bf y}|M_{2}) P(M_{2}) } = B_{12}\cdot\frac{P(M_{1})}{P(M_{2})}.
\label{eq:modelpost}
\end{equation}
The ratio, when the prior probabilities of the models are equal, is reduced to the Bayes factor of Eq. (\ref{eq:B12}). The assessment of the Bayes factors involves two integrals over the parameter space of $({\bm\theta},\sigma^{2})$. The integrands for models $M_{1}$ and $M_{2}$ are $p({\bf y}|{\bm\theta},\sigma^2,M_{1}) \cdot p({\bm\theta},\sigma^{2}|M_{1})$ and $p({\bf y}|{\bm\theta},\sigma^2,M_{2}) \cdot p({\bm\theta},\sigma^{2}|M_{2})$, respectively. 

For general multi-dimensional integration, asymptotic approximation or simulation-based estimation are two commonly-used methods. Laplace approximation method evaluates the integral as a multivariate normal distribution. Consider the above integrand term in Eq. (\ref{eq:B12}), i.e., $p({\bf y}|{\bm\theta},\sigma^2,M_{1}) \cdot p({\bm\theta},\sigma^{2}|M_{1})$. Drop the model symbol for simplicity of the derivation and denote $({\bm\theta},\sigma^2)$ as $\phi$. The natural logarithm of the integrand, $p({\bf y}|\phi)p(\phi)$, i.e., $p({\bf y},\phi)$, can be expressed using Taylor expansion around its mode $\phi'$ as
\begin{equation}
\begin{array}{rl}
  \ln p({\bf y}, \phi) 
& = \ln p({\bf y}, \phi') + (\phi - \phi')^{T}\nabla\ln p({\bf y},\phi') \\
&+ \displaystyle{\frac{1}{2}(\phi-\phi')^{T}\left[ \nabla^{2}\ln p({\bf y},\phi')\right] (\phi-\phi')} \\
&+ O\left[ (\phi-\phi')^3\right],
\end{array}
\label{eq:lap2}
\end{equation}
where $\nabla\ln p({\bf y},\phi')$ is the gradient of $\ln p({\bf y},\phi)$ evaluated at $\phi'$, $\nabla^{2}\ln p({\bf y},\phi')$ is the Hessian matrix of $\ln p({\bf y},\phi)$ evaluated at $\phi'$, and $O\left[\cdot\right]$ are higher order terms. Given that the higher order terms are negligible with respect to the other terms, 
\begin{equation}
\begin{array}{rl}
\ln p({\bf y},\phi) & \approx \ln p({\bf y},\phi')+
\displaystyle{\underbrace{(\phi-\phi')^{T}\nabla\ln p({\bf y},\phi')}_{(\circ)}} \\
&+\displaystyle{\frac{1}{2}(\phi-\phi')^{T}\left[\nabla^{2}\ln p({\bf y},\phi') \right](\phi-\phi')}.
\end{array}
\label{eq:lap3}
\end{equation}
The term $(\circ)$ is zero at the mode of the distribution where the gradient is zero; therefore, expanding $\ln p({\bf y},\phi)$ around $\phi'$ eliminates the term $(\circ)$ and yields
\begin{equation}
\ln p({\bf y},\phi) \approx \ln p({\bf y},\phi')+\frac{1}{2}(\phi-\phi')^{T}\left[\nabla^{2}\ln p({\bf y},\phi') \right](\phi-\phi').
\label{eq:lap4}
\end{equation}
Exponentiation of the above equation to obtain
\begin{equation}
\resizebox{\columnwidth}{!}{$
p({\bf y},\phi) \approx p({\bf y},\phi')\exp\left\{ -\displaystyle\frac{1}{2}\left(\phi-\phi'\right)^{T}
\left[ -\nabla^{2}\ln p({\bf y},\phi') \right] \left(\phi -\phi'\right) \right\}.
$}
\label{eq:pdfapprox}
\end{equation}
Realizing the first term of the above equation is a constant, and the last term is the variable part of a multivariate normal distribution with a mean vector of $\phi'$ and a covariance matrix $\Sigma' = [-\nabla^{2}\ln p({\bf y},\phi')]^{-1}$, the integration of $p({\bf y},\phi)$ writes,
\begin{equation}
\resizebox{1.0\columnwidth}{!}{$
\begin{array}{rl}
P({\bf y}) &= p({\bf y},\phi')\cdot \displaystyle\int_{\Phi}\exp\left[-\frac{1}{2}(\phi-\phi')^{T}\left[\Sigma' \right]^{-1}(\phi-\phi') \right]\mathrm{d}\phi \\
&= p({\bf y},\phi')\sqrt{(2\pi)^{k+1}\lvert\Sigma'\rvert}.
\end{array}
$}
\label{eq:pdfint}
\end{equation}
Notice that ${\bm\theta}$ is a $k$-dimensional vector so $\phi=({\bm\theta},\sigma^{2})$ is a $k+1$ dimensional vector. Using the result of Eq. (\ref{eq:pdfint}), the integral associated with $j$th model in Eq. (\ref{eq:B12}) is
\begin{equation}
P({\bf y}|M_{j}) \approx p\left({\bf y}, ({\bm\theta},\sigma^{2})' |M_{j}\right)\sqrt{(2\pi)^{k+1}\lvert \Lambda_{j}'\rvert}.
\label{eq:modk}
\end{equation}
where $({\bm\theta},\sigma^2)'$ is the mode of the distribution and $\Lambda_{j}'$ is the covariance matrix, i.e., the inverse of the negative Hessian matrix of $\ln p({\bf y},{\bm\theta},\sigma^{2}|M_{j}) $ evaluated at $({\bm\theta},\sigma^{2})'$. Proper numerical derivative methods based on finite difference schemes can achieve reliable results for the gradient and the Hessian in the following equation. 
\begin{equation}
\left\{
\begin{aligned}
	({\bm\theta}, \sigma^2)' &= \underset{({\bm\theta}, \sigma^2)}{\mathop{\arg\max}}\left[ \nabla\ln p({\bf y},{\bm\theta},\sigma^{2}|M_{j}) \right]  \\
	\Lambda' &= \left[-\nabla^{2} \ln p({\bf y},{\bm\theta},\sigma^{2}|M_{j})  \biggr\rvert_{({\bm\theta}, \sigma^2)'} \right]^{-1}
	\end{aligned}
		\right.,
\label{eq:lapsol}
\end{equation}
Experiences have shown that for a distribution with a single mode and is approximately symmetric, the Laplace approximation method can achieve accurate results for engineering applications \cite{guan2012efficient,he2016improve}. 

It should be noted that $1/\sigma^{q}$-type of noninformative prior is not a proper prior, i.e., the normalizing constant for $1/\sigma^{q}$ is not bounded for $\sigma\in(0,+\infty)$; therefore, it is necessary to limit the support of $1/\sigma^{q}$ to a proper range and to have a finite normalizing constant. Assume the effective range of $\sigma$ is $(\sigma_{-},\sigma_{+})$ for $q \in (0,1,...)$, the normalizing constant is 
\begin{equation}
Z(\sigma) = \left\{
\begin{aligned}
\ln(\sigma_{+1})-\ln(\sigma_{-}) \quad\quad & q = 1 \\
\frac{1}{1-q}\left(\sigma_{+}^{1-q} - \sigma_{-}^{1-q} \right) \quad\quad & q \neq 1
\end{aligned}\right. .
\label{eq:zsig}
\end{equation}
Substitute $p(\bm\theta,\sigma^{2})$ with the $1/(Z(\sigma)\sigma^{q})$ in Eq. (\ref{eq:bayesian}). The global likelihood of Eq. (\ref{eq:modk}) compares the fitting performance of different $1/\sigma^{q}$-type of priors. 
\subsection{Predictive performance}
The predictive performance for a model is measured by the global likelihood of the data that are not used for model parameter estimation. The global likelihood of the future measurement data $\tilde{{\bf y}}$ with the posterior PDF of $p({\bm\theta},\sigma^2|{\bm y})$ writes
\begin{equation}
\begin{aligned}
P(\tilde{{\bf y}}) & = \int p(\tilde{{\bf y}}|{\bm\theta},\sigma^2) p({\bm\theta},\sigma^2|{\bf y}) \mathrm{d}{\bm\theta}\mathrm{d}\sigma^2 \\
& = \frac{1}{P({\bf y})}\int p(\tilde{{\bf y}}|{\bm\theta},\sigma^{2}) p({\bf y}|{\bm\theta},\sigma^2)p({\bm\theta},\sigma^2)\mathrm{d}{\bm\theta}\mathrm{d}\sigma^{2},
\end{aligned}
\label{eq:logpred}
\end{equation}
where $P({\bf y})$ is evaluated using Eq. (\ref{eq:modk}). 

The comprehensive performance integrating both fitting performance and predictive performance in terms of the global likelihood in the joint space of $({\bf y}\times\tilde{{\bf y}})$ is 
\begin{equation}
P(\tilde{{\bf y}})P({\bf y}) = \int p(\tilde{{\bf y}}|{\bm\theta},\sigma^{2})p({\bf y}|{\bm\theta},\sigma^{2})p({\bm\theta},\sigma^2)\mathrm{d}{\bm\theta}\mathrm{d}\sigma^{2}.
\label{eq:ptot}
\end{equation}
Realize that Eq. (\ref{eq:pdfint}), Eq. (\ref{eq:logpred}) and Eq. (\ref{eq:ptot}) can all be evaluated using the method of Laplace approximation, and the performance of different priors can quantitatively be compared. 

\section{Application examples}
To demonstrate the proposed Bayesian inference with noninformative priors, several examples are presented. Results are compared with the classical least square approach. The fitting and comprehensive performance of $1/\sigma^{q}$-type of priors are evaluated using the method of Laplace approximation.
\subsection{A simple linear regression problem}
\begin{figure*}[!ht]
\centering
\subfloat[]{\label{fig:prior_case_3b}\includegraphics[width=0.4\textwidth]{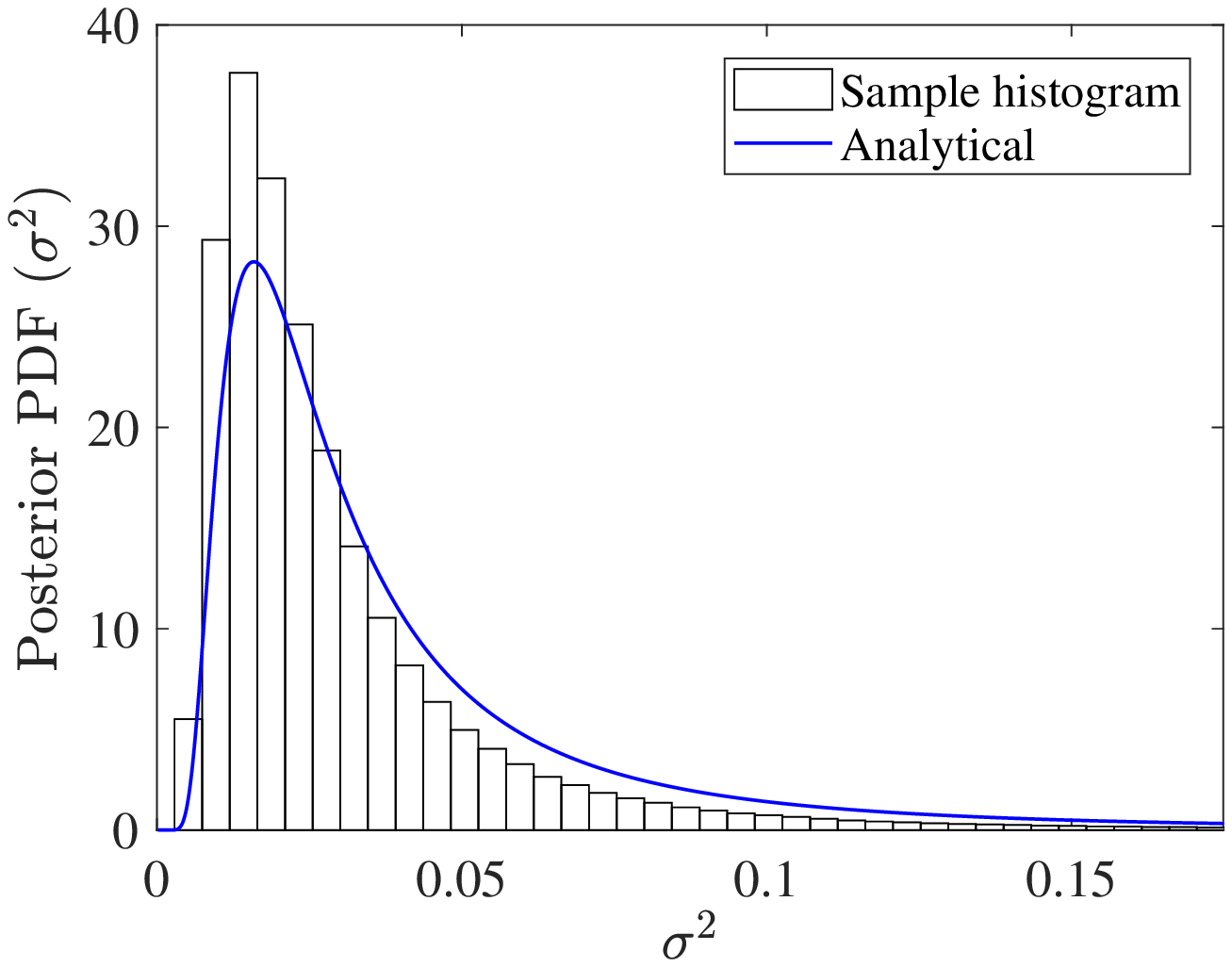}} 
\subfloat[]{\label{fig:prior_case_3c}\includegraphics[width=0.4\textwidth]{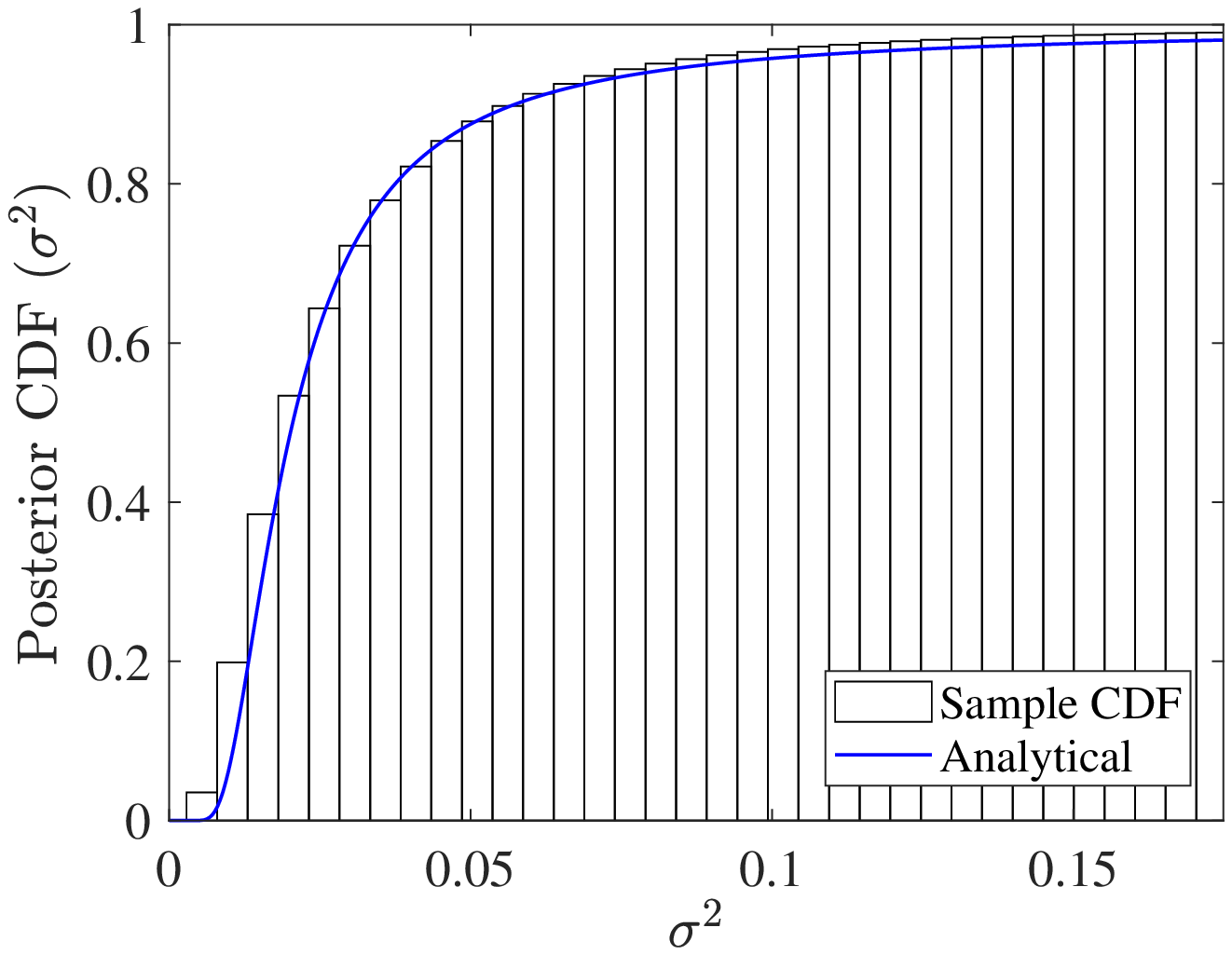}} \\ [+4ex]
\subfloat[]{\label{fig:prior_case_3a}\includegraphics[width=0.4\textwidth]{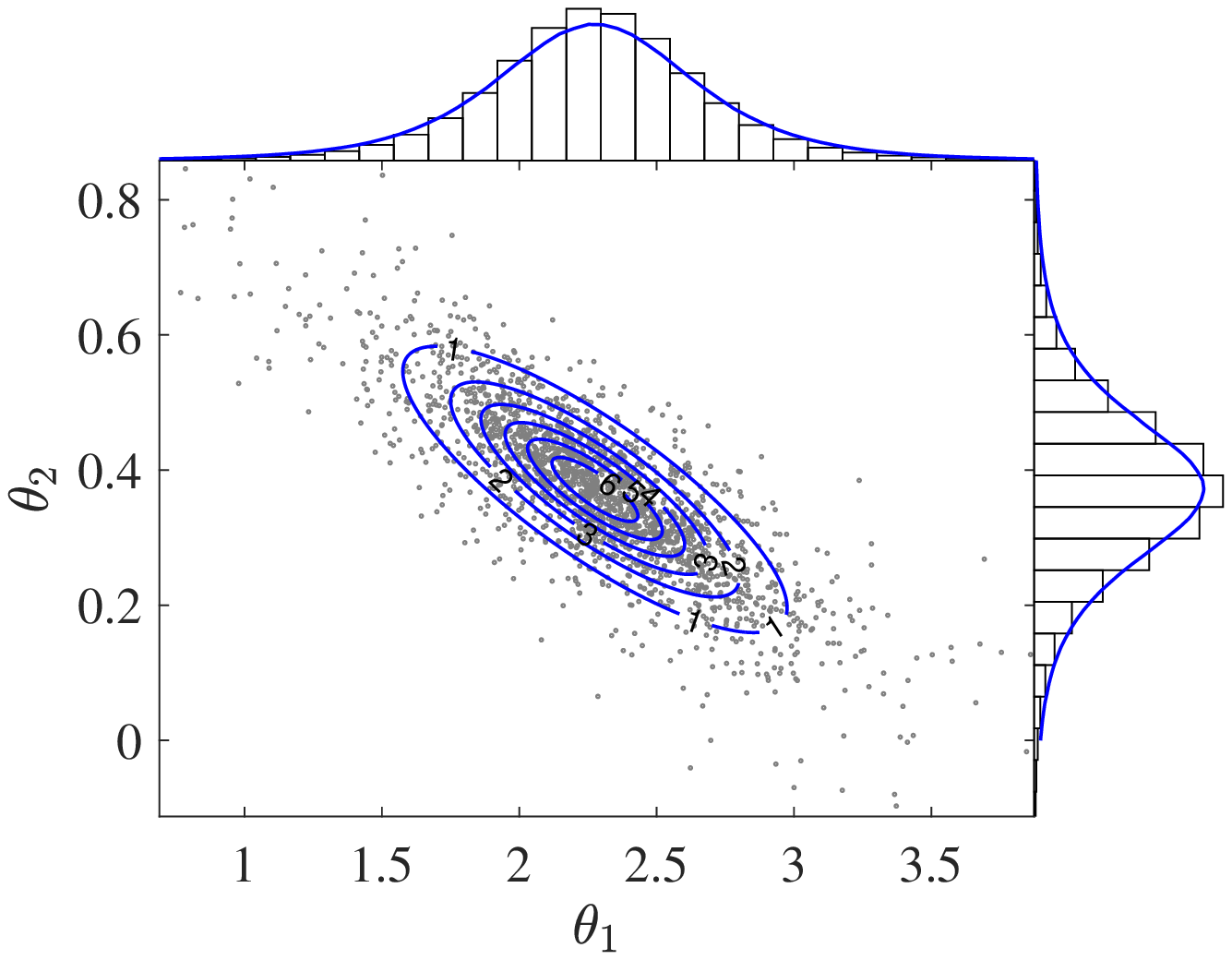}}\quad\quad
\subfloat[]{\label{fig:prior_case_3d}\includegraphics[width=0.4\textwidth]{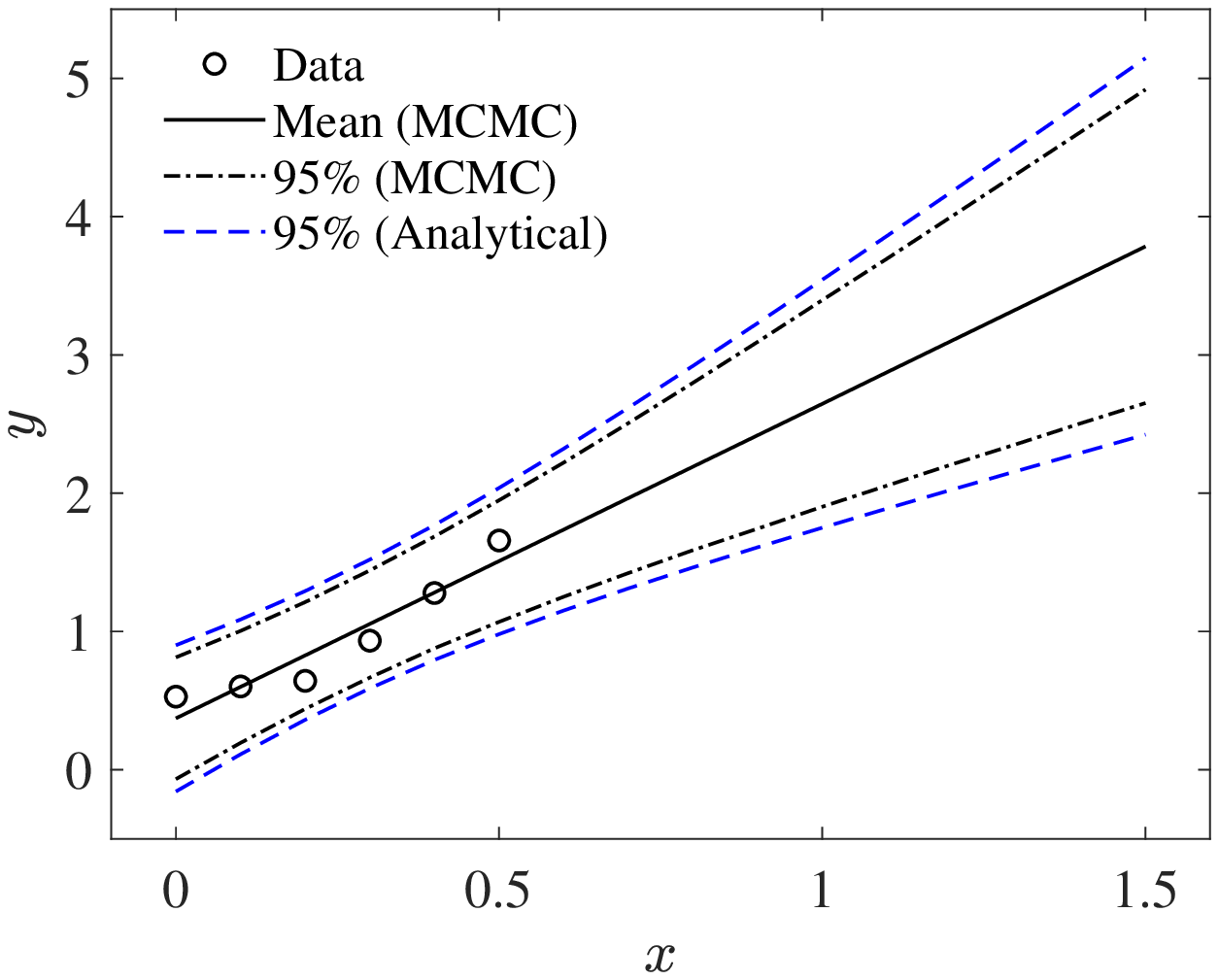}}
\caption{(a) Histogram of MCMC samples of $\sigma^2$, (b) empirical CDF of $\sigma^2$, (c) samples and histogram of ${\bm\theta}$, and (d) mean prediction and bounds. The solid lines in (a)--(c) represent theoretical values by Eq. (\ref{eq:postsigma2theory}) and Eq. (\ref{eq:postthetatheory}).}%
\label{fig:prior_case_n6p2_1to4}%
\end{figure*}

A simple linear regression problem is used to demonstrate the idea of noninformative Bayesian regression. Ten points of data are generated $y=3x+0.25+0.3\epsilon$, where $\epsilon\sim \mathrm{Norm}(0,1)$, as shown in Table \ref{tab:linearexp}. 
\begin{table*}[h]%
\centering
\caption{Data of the linear regression example}
\begin{tabular}{llllllllllll}
Data & 1 & 2 & 3 & 4 & 5 & 6 &7 &8 &9 & 10 \\
\hline
x & 0 & 0.1 & 0.2 & 0.3 & 0.4 & 0.5 & 0.6 & 0.7 & 0.8 & 0.9 \\
y & 0.2784 & 0.3520 & 0.3925 & 0.6831 & 1.0277 & 1.4077 & 1.8802 & 2.5003 & 2.0007 & 3.1197 \\ 
\hline
\end{tabular}
\label{tab:linearexp}
\end{table*}
A Gaussian likelihood with an unknown scale parameter is used for illustration, the joint posterior of $({\bm\theta},\sigma^2)$ is
\begin{equation}
p({\bm\theta},\sigma^2|\mathbf{y})\propto p(\bm\theta,\sigma^2)\cdot \left( \frac{1}{\sigma^{2}} \right)^{n/2} \exp\left[-\frac{\sum_{i}^{n}\left(y_{i}-{\bf{x}}_{i}^{T}{\bm\theta} \right)^2}{2\sigma^2} \right],
\label{eq:appeq1}
\end{equation}
where the first term of the right hand side of Eq. (\ref{eq:appeq1}) is the undetermined prior of $(\bm\theta,\sigma^2)$. Using the Jeffreys prior, i.e., $p(\bm{\theta},\sigma^2)\propto 1/\sigma^2$, the posterior writes
\begin{equation}
p({\bm\theta},\sigma^2|\mathbf{y})\propto \left( \frac{1}{\sigma^{2}} \right)^{1+n/2}\exp\left[-\frac{\sum_{i}^{n}\left(y_{i}-{\bf{x}}_{i}^{T}{\bm\theta} \right)^2}{2\sigma^2} \right].
\label{eq:appeq2}
\end{equation}
For illustration purposes, the first 6 points are used to obtain the posterior PDF, i.e., $n=6$. A total number of $4\times 10^{6}$ samples are drawn using MCMC simulations with first 2000 burn-in samples dropped. The resulting histograms of the joint distribution are shown in Fig. \ref{fig:prior_case_n6p2_1to4}. 
The sampling method can be avoided when consider the Jeffreys prior as a limiting state of conjugate NIG prior. Consequently, using Eq. (\ref{eq:postsigma2theory}) and Eq. (\ref{eq:postthetatheory}) the posterior PDF of $\sigma^2$ and $\bm{\theta}$ can analytically be obtained with $n=6$ and $k=2$. The asymptotic analytical solution has the advantage when the evaluation is subject to time constraint. The difference between the MCMC sampling and the analytical solution is presented in Fig. \ref{fig:prior_case_n6p2_1to4} where a close agreement of the two methods is observed.

\subsubsection{Comparison with regular least square}
To signify the difference between the noninformative Bayesian (NB) linear regression and regular least square (LS) linear regression, the model predictions of the two methods are visually compared as shown in Fig. \ref{fig:prior_case_n6p2_1to4}. Results are obtained with 6 data points. It is observed that using the same data points for regression, noninformative Bayesian regression yields narrower confidence bounds than that of regular least square regression. From the statistical point of view, noninformative Bayesian regression gives better performance in terms of uncertainties. 

\begin{figure*}
\centering
\subfloat[]{\label{fig:prior_case_n4p2_5}\includegraphics[width=0.4\textwidth]{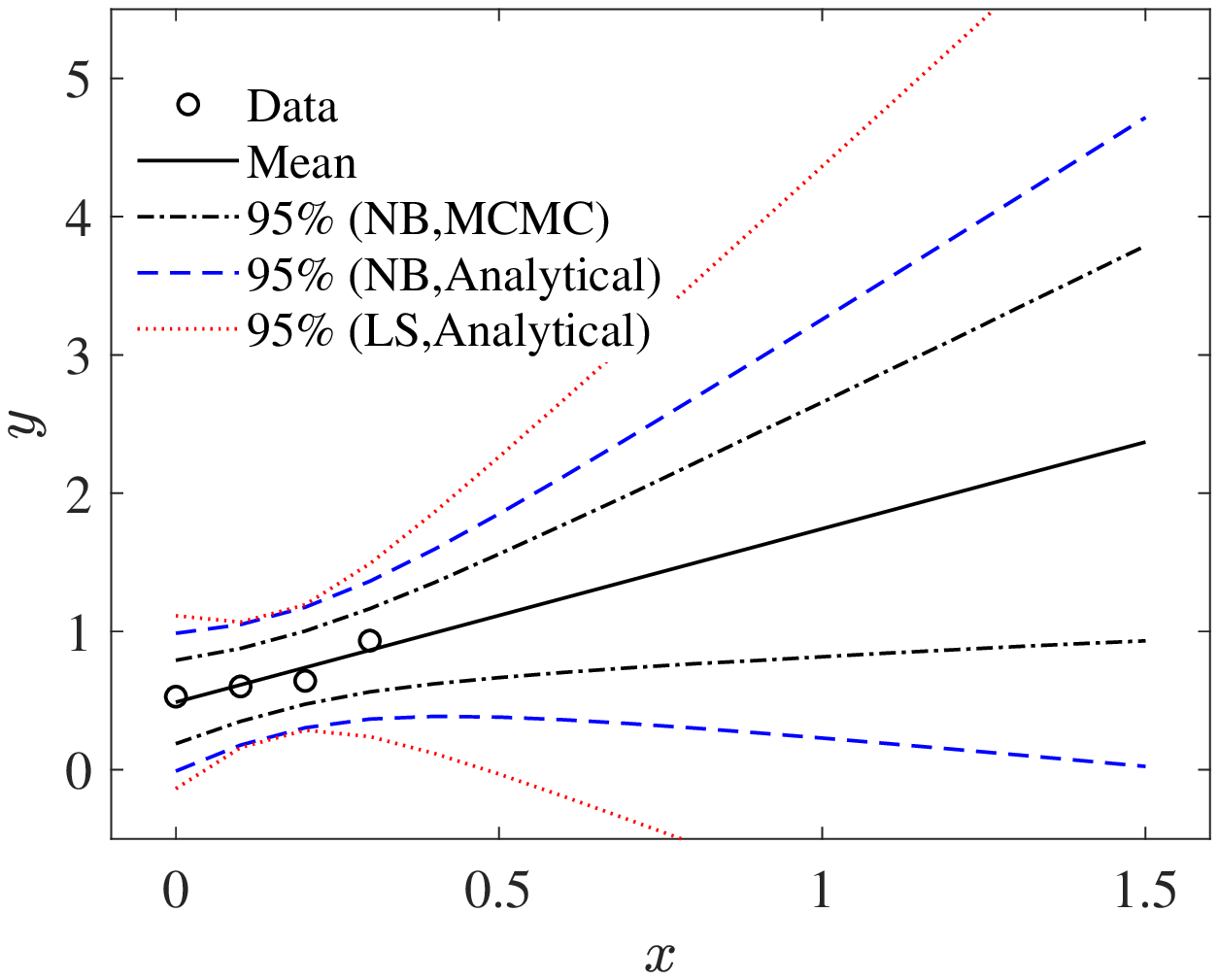}}\quad\quad
\subfloat[]{\label{fig:prior_case_n5p2_5}\includegraphics[width=0.4\textwidth]{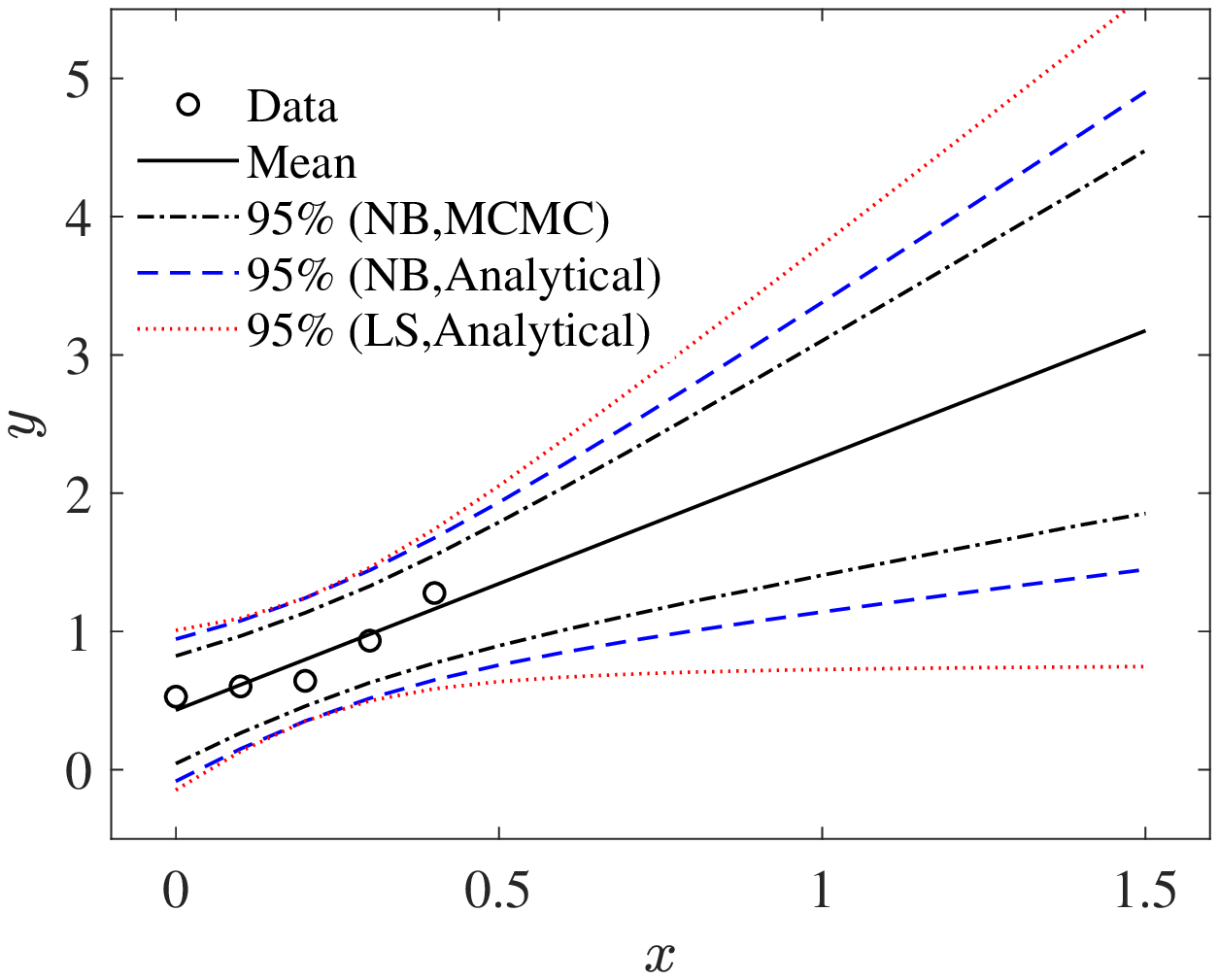}}\\
\subfloat[]{\label{fig:prior_case_n6p2_5}\includegraphics[width=0.4\textwidth]{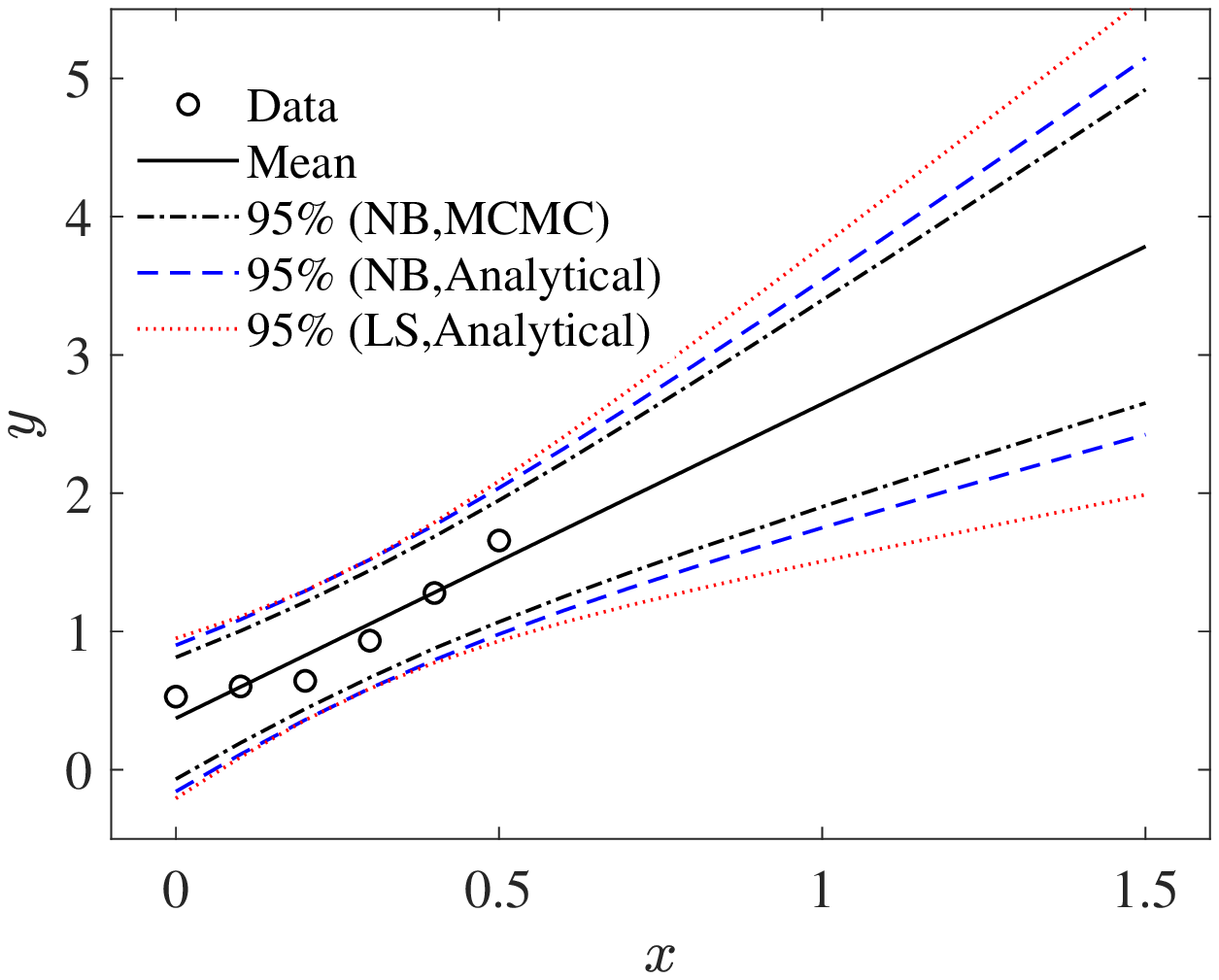}}\quad\quad
\subfloat[]{\label{fig:prior_case_n7p2_5}\includegraphics[width=0.4\textwidth]{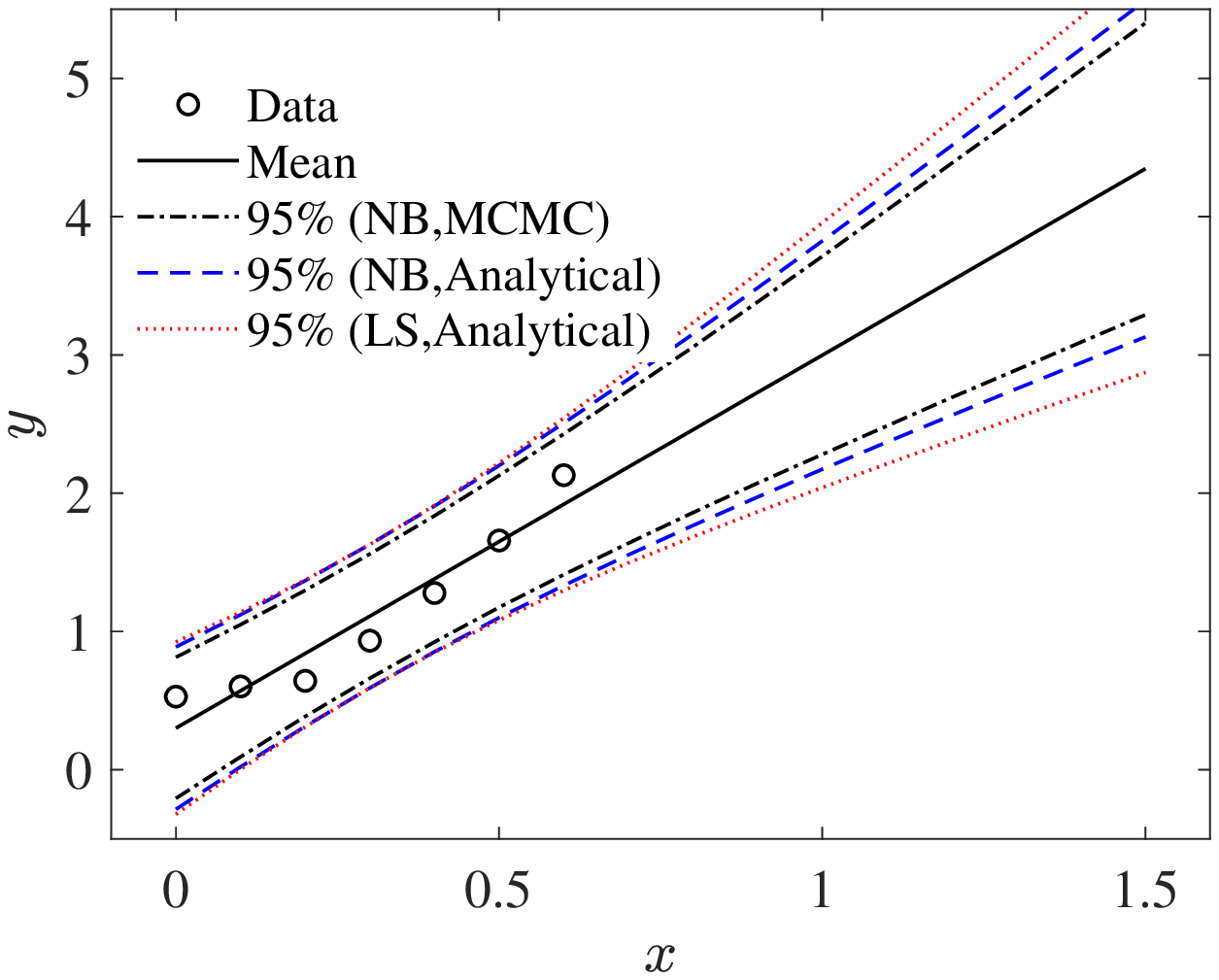}}\\
\subfloat[]{\label{fig:prior_case_n8p2_5}\includegraphics[width=0.4\textwidth]{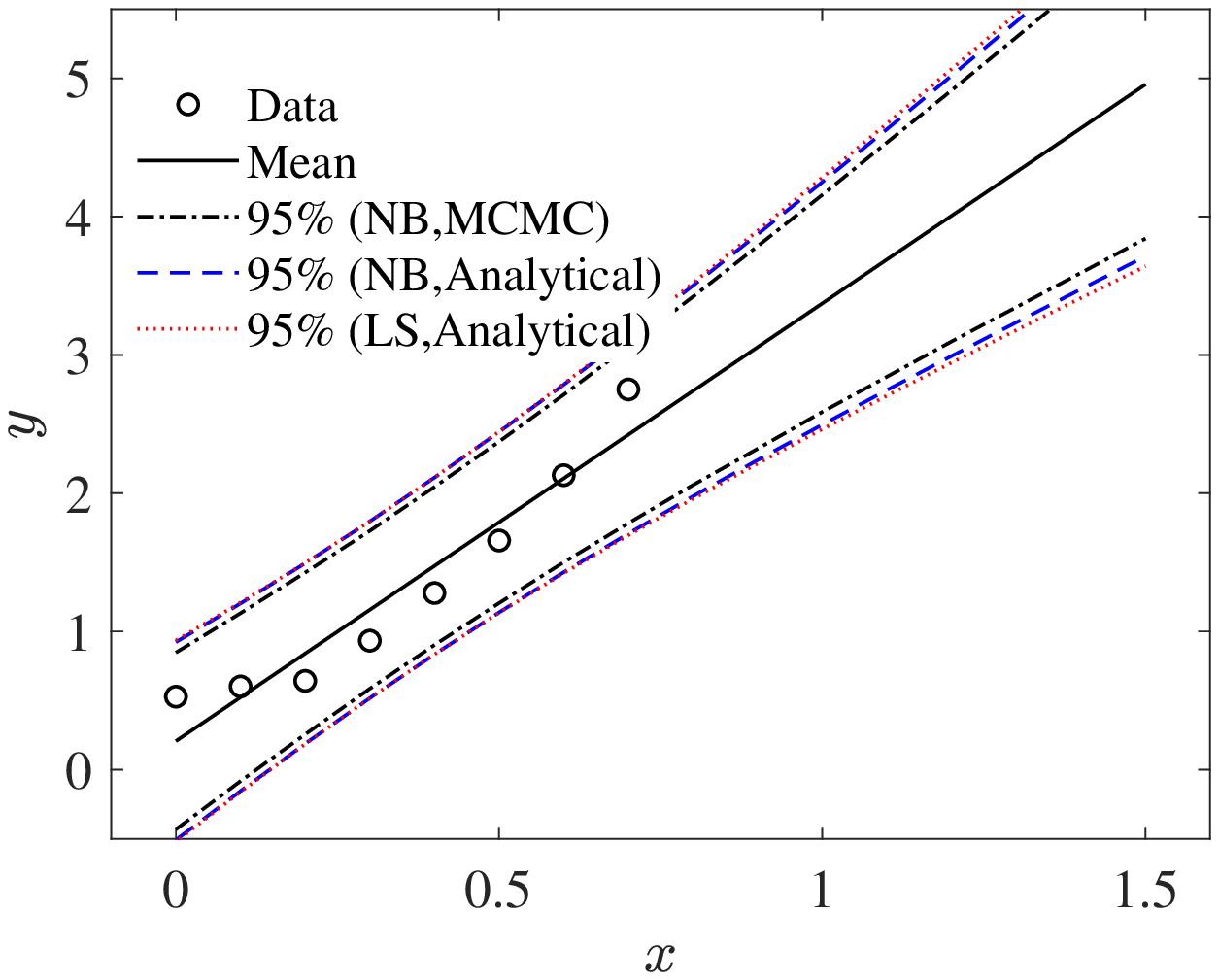}}\quad\quad
\subfloat[]{\label{fig:prior_case_n9p2_5}\includegraphics[width=0.4\textwidth]{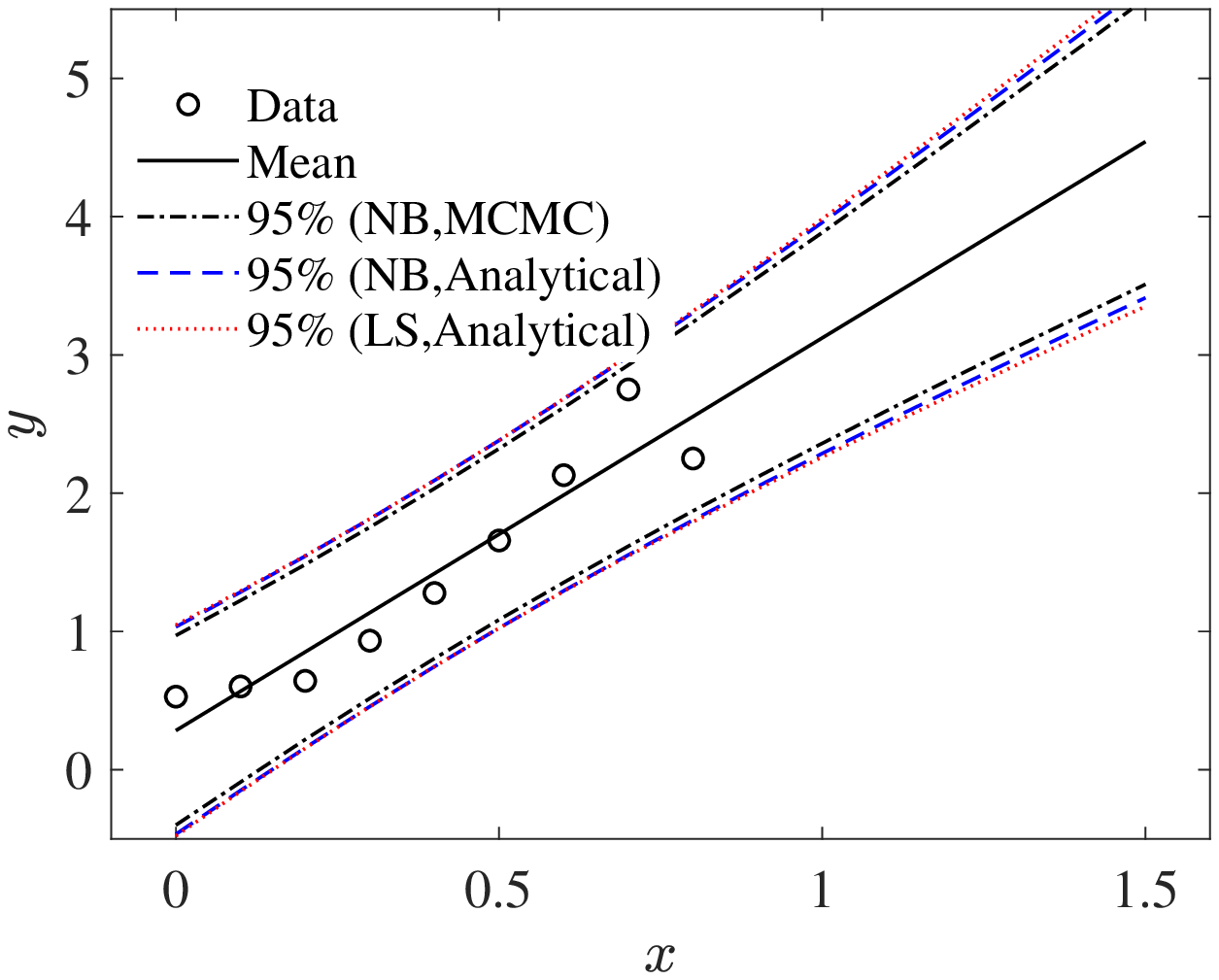}}
\caption{Comparisons of noninformative Bayesian regression and least square regression on prediction intervals. (a) 4 data points, (b) 5 data points, (c) 6 data points, (d) 7 data points, (e) 8 data points, and (f) 9 data points.}%
\label{fig:prior_case_n4to9p2_5}%
\end{figure*}

It is known that when the number of data points used for regression increases, the resulting Student's {\it t}--distribution of the model parameters gradually converges to a normal distribution. The difference between the two methods becomes smaller as more data points are used. To verify this, additional numerical studies are made with an increasing number of data points. Fig. \ref{fig:prior_case_n4to9p2_5} presents the comparison results of noninformative Bayesian regression and the least square regression. It can be seen that the difference between the prediction intervals of the two is large when limited data points are used, and the difference becomes smaller as the number of data points increases. For example, when using 9 data points the prediction intervals only show a slight difference, as shown in Fig. \ref{fig:prior_case_n4to9p2_5}(f).

\subsubsection{Comparisons on priors}
The overall performance consists of fitting performance and prediction performance, both of which can be measured using the global likelihood of the data. Asymptotic results of the posterior of model parameters and predictions using different $1/\sigma^{q}$ priors are obtained and shown in Table \ref{tab:compare_prior}. MCMC samples are also drawn from the posteriors for comparisons.
\begin{table*}%
\centering
\caption{$1/\sigma^{q}$ type of priors for $(\theta,\sigma^2)$, its Bayesian posteriors, and model predictions. The histograms and scatter points in the second column are results from $4\times 10^6$ MCMC simulations; the solid line in the second column are results of the analytical forms of the posteriors. The third column presents the model predictions, where the circle represents the data points for regression, the dash-dot lines shows the 95\% prediction intervals based on the MCMC samples of the posterior distribution, and the dash lines shows the 95\% prediction intervals based on the analytical forms of the posterior distribution.}
\begin{tabular}{lcc}
Prior $(\theta,\sigma^2)$ & Posterior $(\theta,\sigma^2|y)$ & Model prediction $(\tilde{y}|y)$ \\
\hline
flat 
& \hspace{0.2cm}\begin{minipage}{.5\textwidth}\vspace{0.5cm}\includegraphics[width=.7\linewidth]{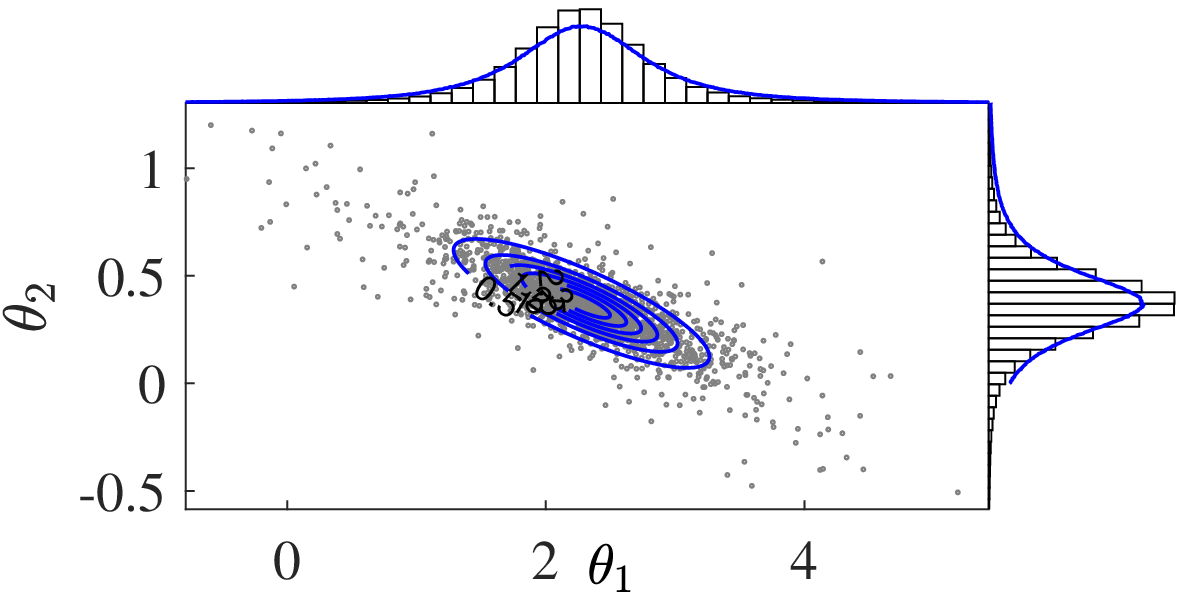}\includegraphics[width=.3\linewidth]{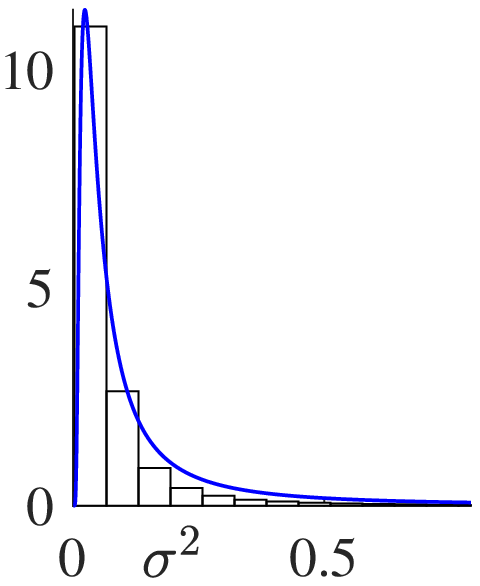}\end{minipage}\hspace{0.0cm} 
& \begin{minipage}{.275\textwidth}\vspace{0.5cm}\includegraphics[width=\linewidth]{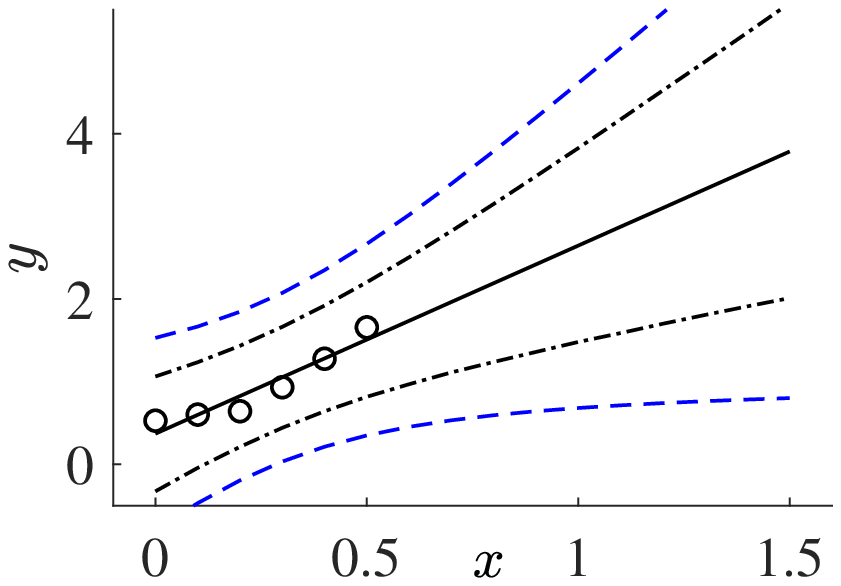}\end{minipage} \\
$\frac{1}{\sigma}$ 
& \hspace{0.2cm}\begin{minipage}{.5\textwidth}\vspace{0.25cm}\includegraphics[width=.7\linewidth]{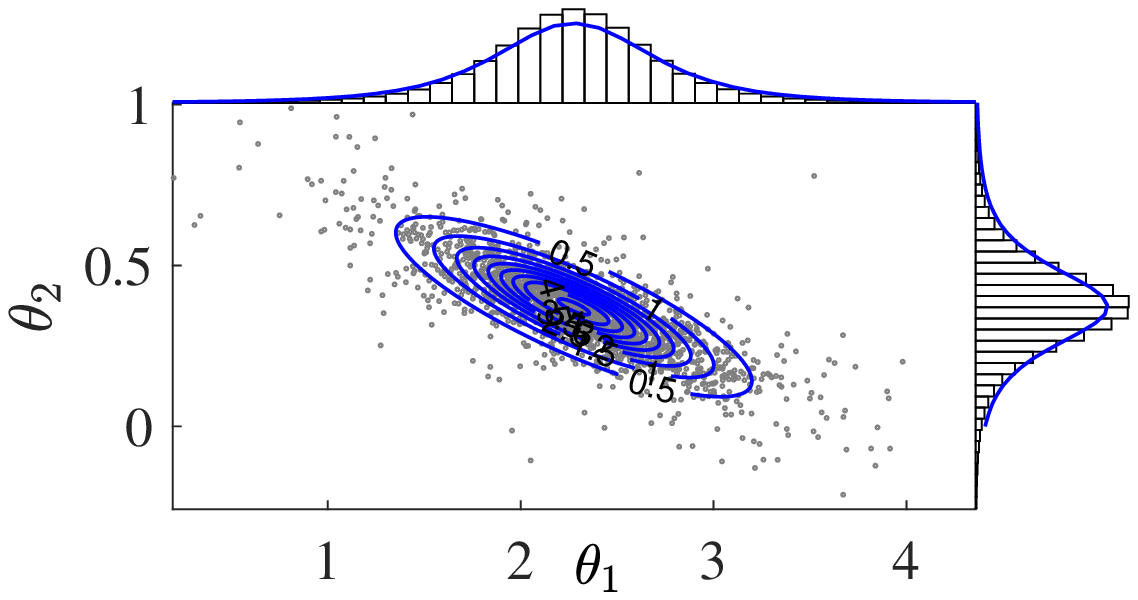}\includegraphics[width=.3\linewidth]{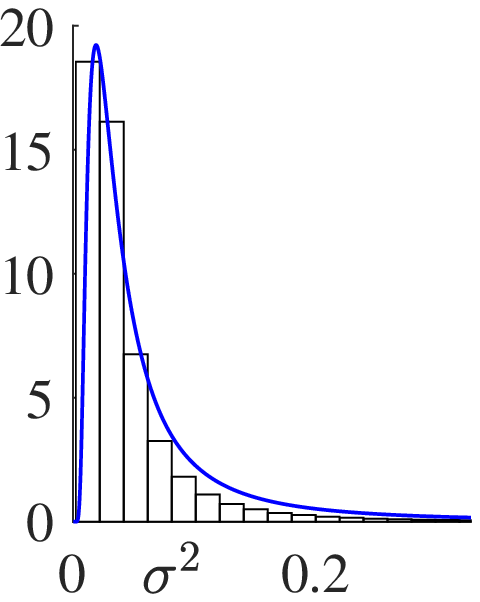}\end{minipage}\hspace{0.0cm} 
& \begin{minipage}{.275\textwidth}\vspace{0.25cm}\includegraphics[width=\linewidth]{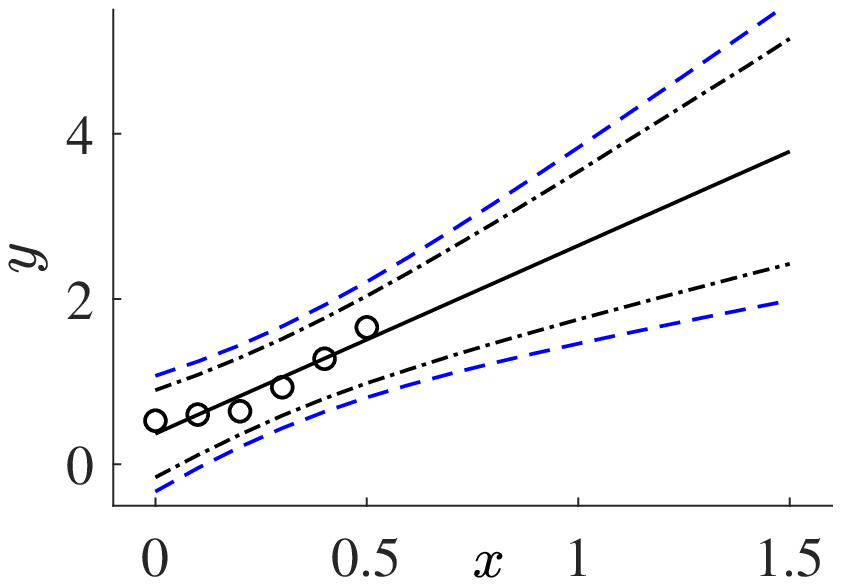}\end{minipage} \\
$\frac{1}{\sigma^2}$ 
& \hspace{0.2cm}\begin{minipage}{.5\textwidth}\vspace{0.25cm}\includegraphics[width=.7\linewidth]{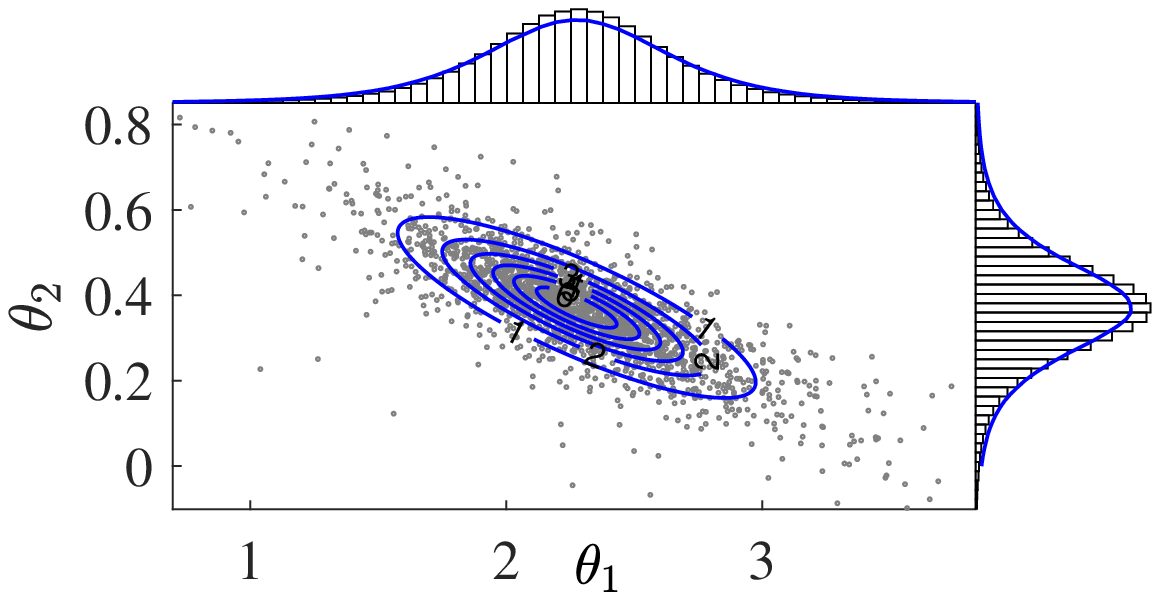}\includegraphics[width=.3\linewidth]{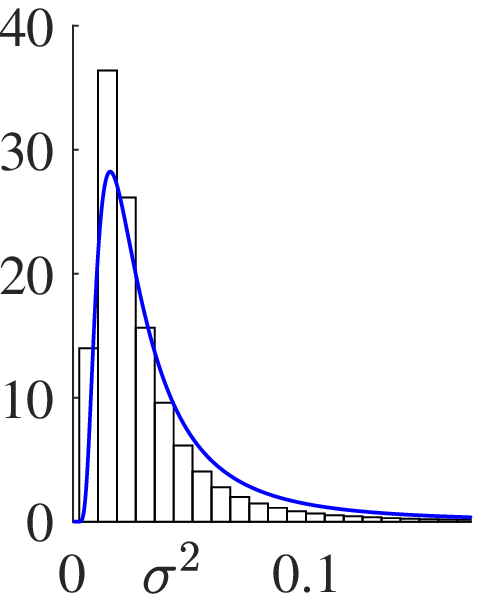}\end{minipage}\hspace{0.0cm} 
& \begin{minipage}{.275\textwidth}\vspace{0.25cm}\includegraphics[width=\linewidth]{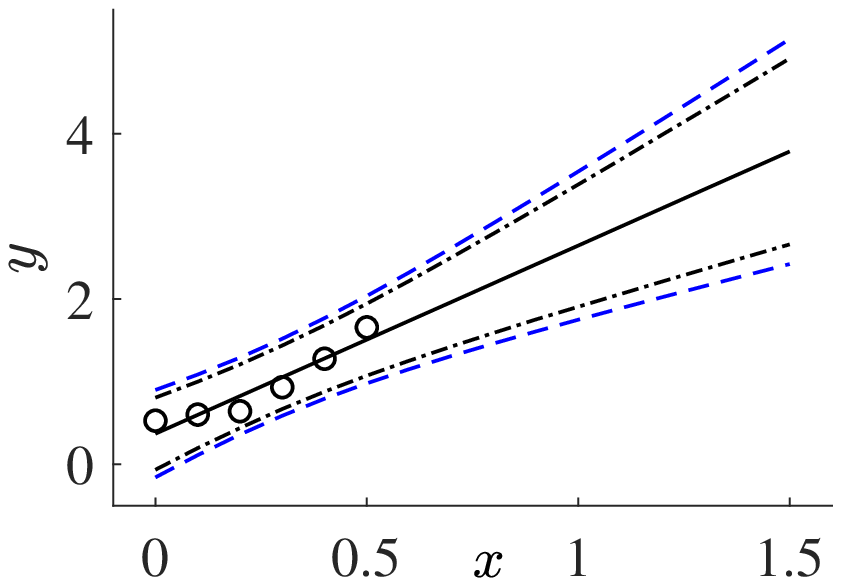}\end{minipage} \\
$\frac{1}{\sigma^3}$ 
& \hspace{0.2cm}\begin{minipage}{.5\textwidth}\vspace{0.25cm}\includegraphics[width=.7\linewidth]{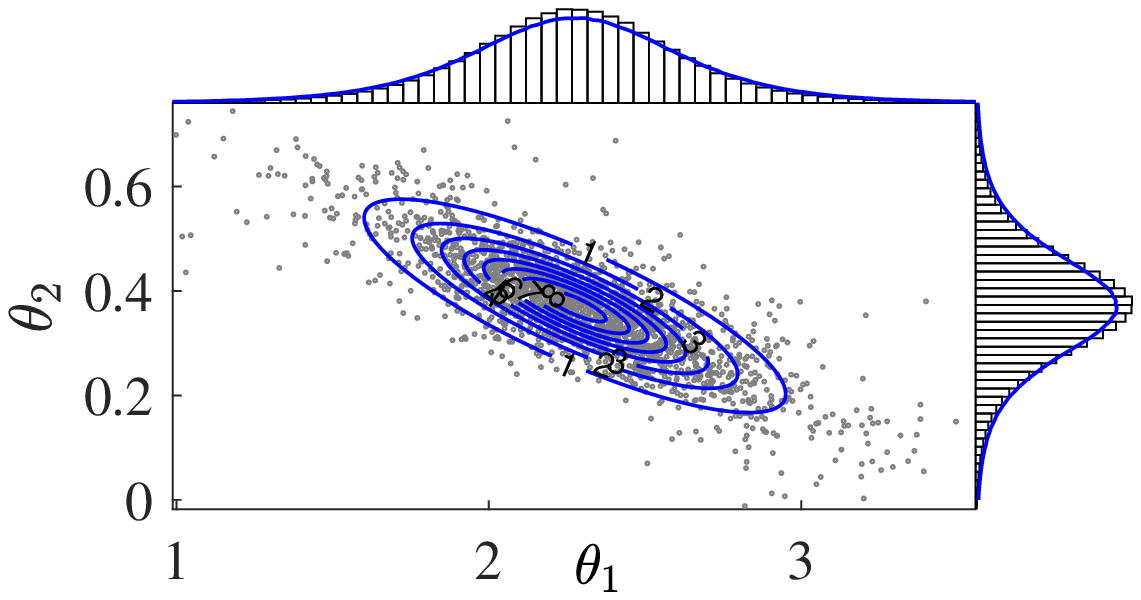}\includegraphics[width=.3\linewidth]{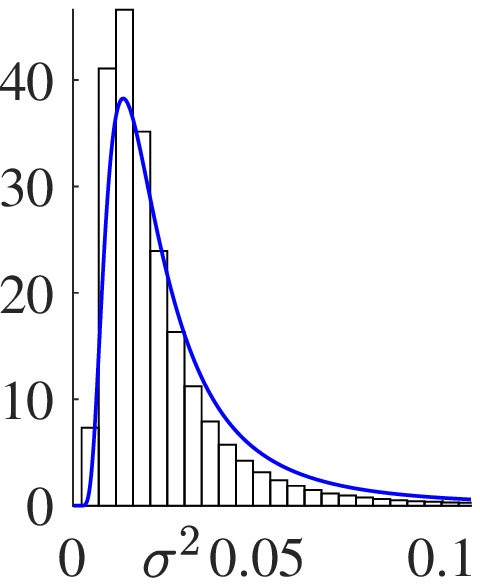}\end{minipage}\hspace{0.0cm} 
& \begin{minipage}{.275\textwidth}\vspace{0.25cm}\includegraphics[width=\linewidth]{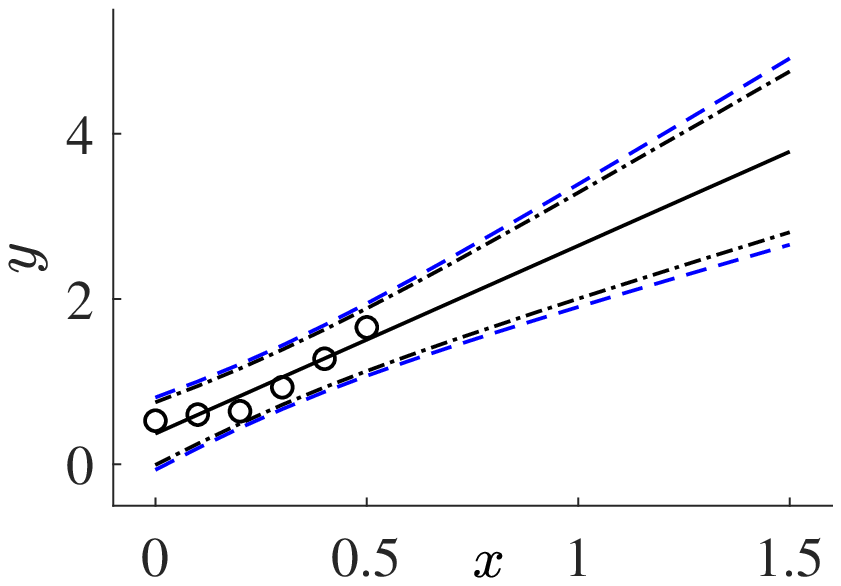}\end{minipage} \\
$\frac{1}{\sigma^4}$ 
& \hspace{0.2cm}\begin{minipage}{.5\textwidth}\vspace{0.25cm}\includegraphics[width=.7\linewidth]{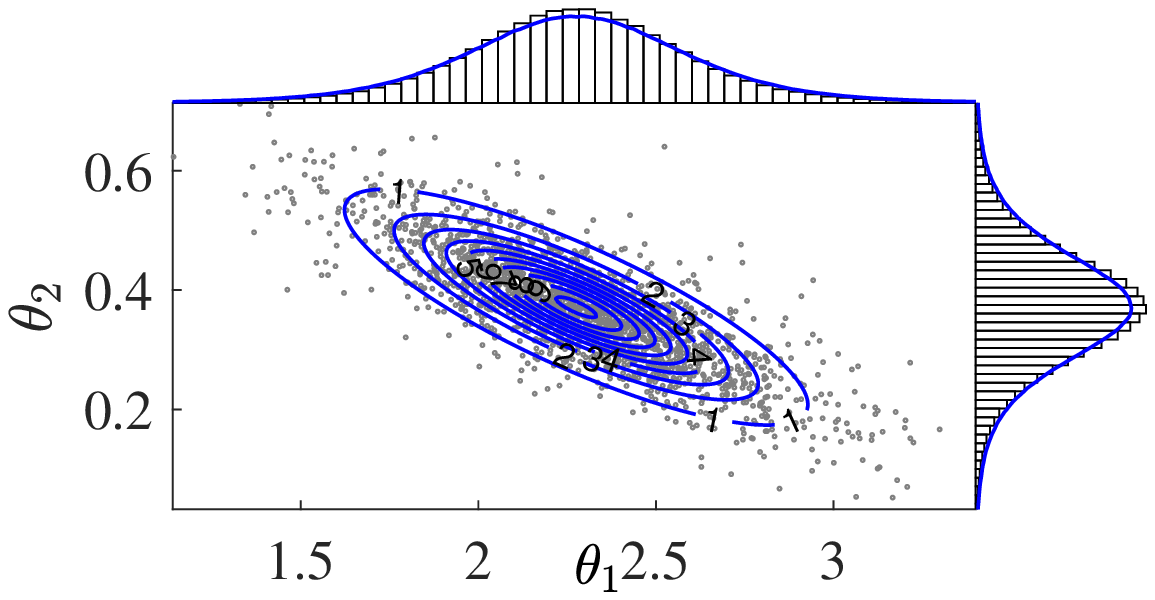}\includegraphics[width=.3\linewidth]{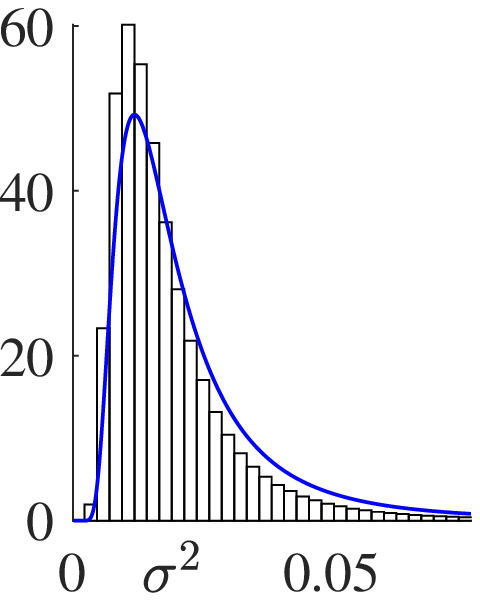}\end{minipage}\hspace{0.0cm} 
& \begin{minipage}{.275\textwidth}\vspace{0.25cm}\includegraphics[width=\linewidth]{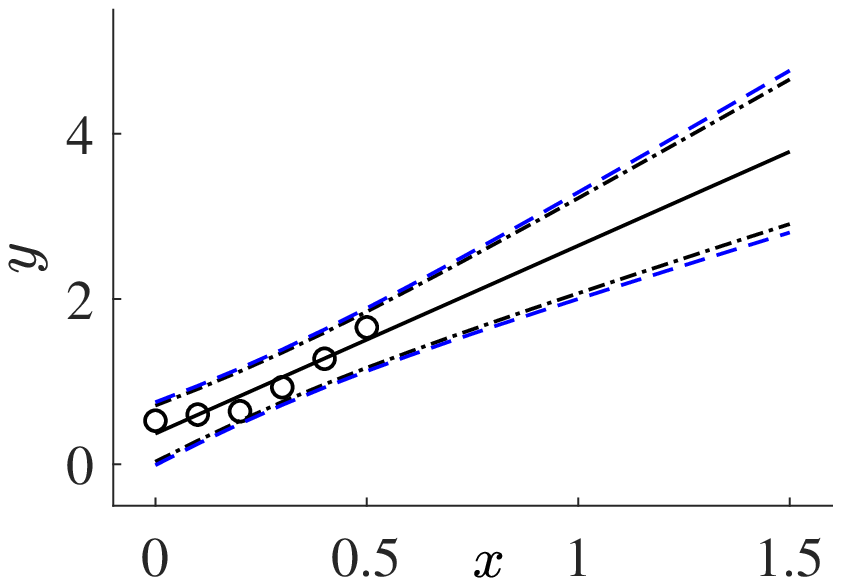}\end{minipage} \\
$\frac{1}{\sigma^5}$ 
& \hspace{0.2cm}\begin{minipage}{.5\textwidth}\vspace{0.25cm}\includegraphics[width=.7\linewidth]{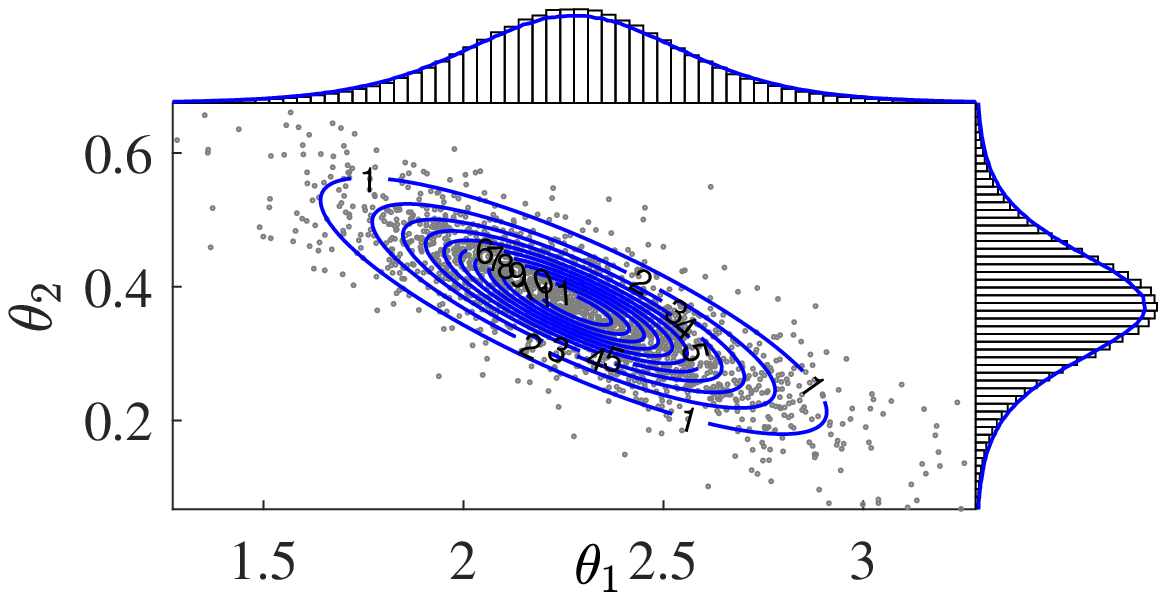}\includegraphics[width=.3\linewidth]{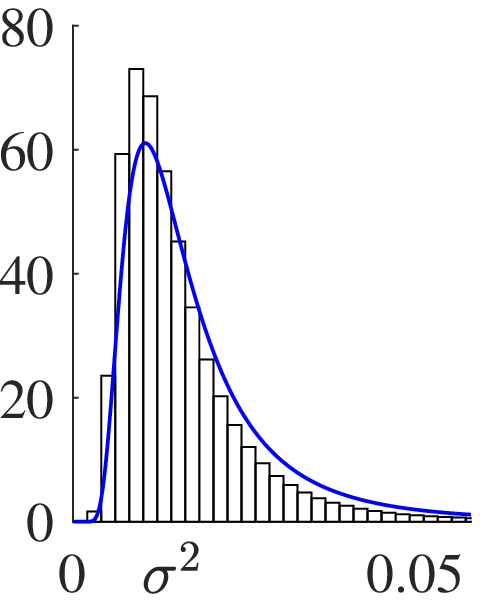}\end{minipage}\hspace{0.0cm} 
& \begin{minipage}{.275\textwidth}\vspace{0.25cm}\includegraphics[width=\linewidth]{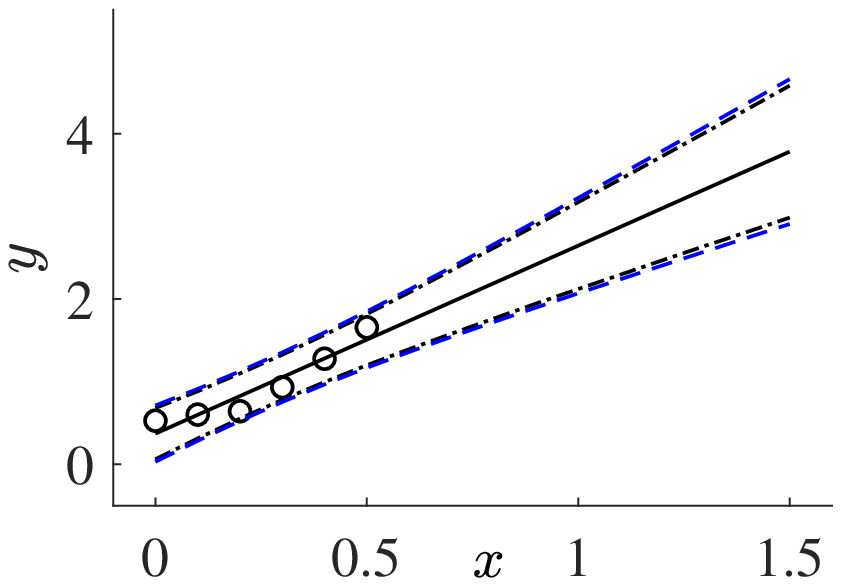}\end{minipage} \\
\vspace{0.1cm}
\end{tabular}
\label{tab:compare_prior}
\end{table*}
It can be seen in figures shown in the third column of Table \ref{tab:compare_prior}, that the 95\% confidence intervals become narrower as $q$ increases. However, this does not necessarily mean a larger $q$ is optimal. 

\begin{figure*}[!ht]
\centering
\subfloat[]{\label{fig:plot_logpy_p_n6_1}\includegraphics[width=0.4\textwidth]{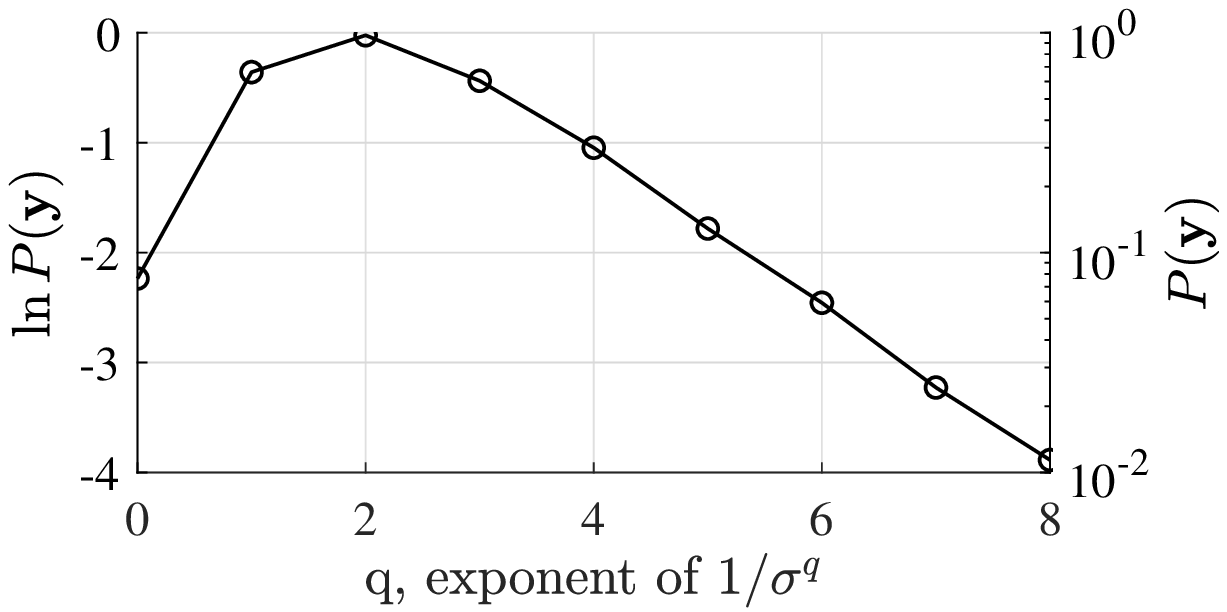}} \quad\quad
\subfloat[]{\label{fig:plot_logpy_p_n6p1_2}\includegraphics[width=0.4\textwidth]{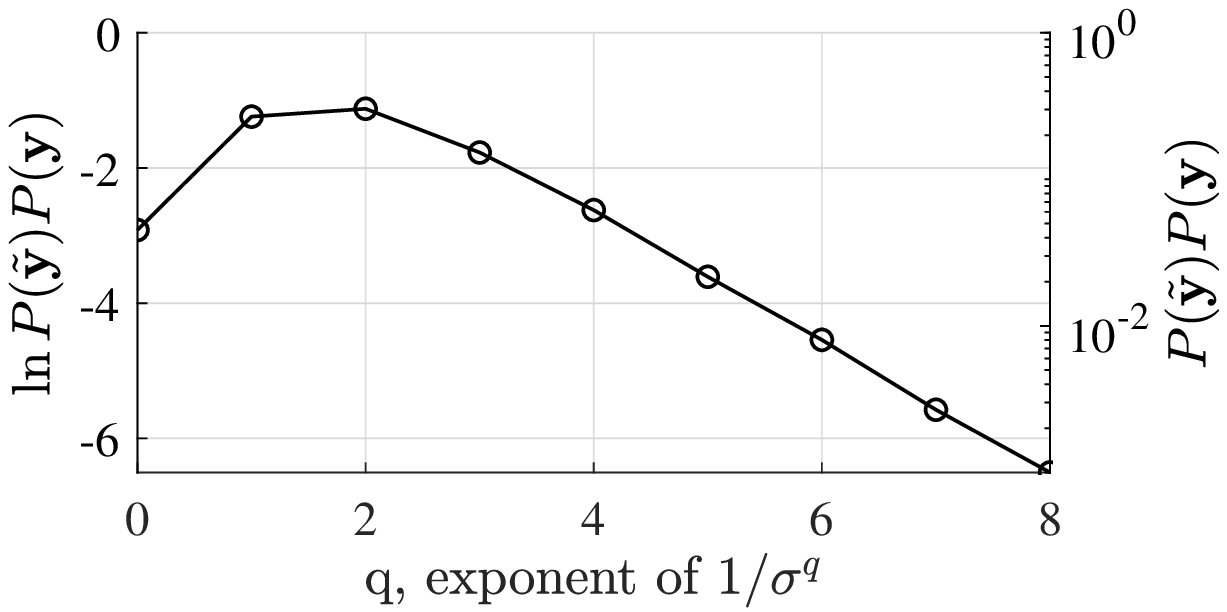}} \\
\caption{(a) $P({\bm y})$ results evaluated using Laplace approximation with different $1/\sigma^{q}$ priors, and (b) $P(\tilde{{\bf y}})P({\bf y})$ results evaluated using Laplace approximation with different $1/\sigma^{q}$ priors. One future data point $\tilde{{\bf y}}$ is considered for prediction.}%
\label{fig:plot_logpy_p_n6}%
\end{figure*}

To reveal which prior performs the fitting better in terms of the global likelihood $P({\bm y})$, results of $P({\bm y})$ using different $1/\sigma^{q}$--type of priors are obtained using numerical methods and are shown in Fig. \ref{fig:plot_logpy_p_n6}(a). In this numerical case the prior of $1/\sigma^{2}$, i.e., $q=2$, gives the maximum global likelihood of data ${\bm y}$. The method of Laplace approximation mentioned earlier is used to evaluate Eq. (\ref{eq:logpred}). One future data point is used and the results with different $q$ are presented in Fig. \ref{fig:plot_logpy_p_n6}(b). Results indicate that $q=2$ yields the maximum likelihood.

To investigate the prior performance for prediction performance. More detailed numerical experiments are conducted. Different number of data points used for estimation and different number of future data points for prediction are combined. The results are summarized in Table \ref{tab:prior_predictive_performance}. Each column in the table represents a particular number of data points ($n$) for parameter estimation, and each row represents the number of future data points ($m$) used for prediction. In particular, results of $P(\bm y)$ associated with $q=0$ are those of regular least square linear regression. It is observed that in this example $1/\sigma$ and $1/\sigma^2$ outperform other priors in all combinations of $m$ and $n$. 
\begin{table*}%
\centering
\caption{The predictive performance of $1/\sigma^{q}$ type of priors for $(\theta,\sigma^2)$. $n$ is the number of points for model parameter estimations, $m$ is the number of future points for model predictions. The $x$-axis is the exponent $q$ of $1/\sigma^{q}$, and the $y$-axis is the global likelihood. The row of $m=0$ is the global likelihood of fitting data, i.e., the fitting performance. Other rows are the global likelihood of a future data given the fitting data, i.e., the overall performance.}
\begin{tabular}{lccc}
\backslashbox[12mm]{m}{n} & 6 & 7 & 8 \\
\hline
\vspace{0.0cm}
0 
& \hspace{0.0cm}\begin{minipage}{.275\textwidth}\vspace{0.5cm}\includegraphics[width=\linewidth]{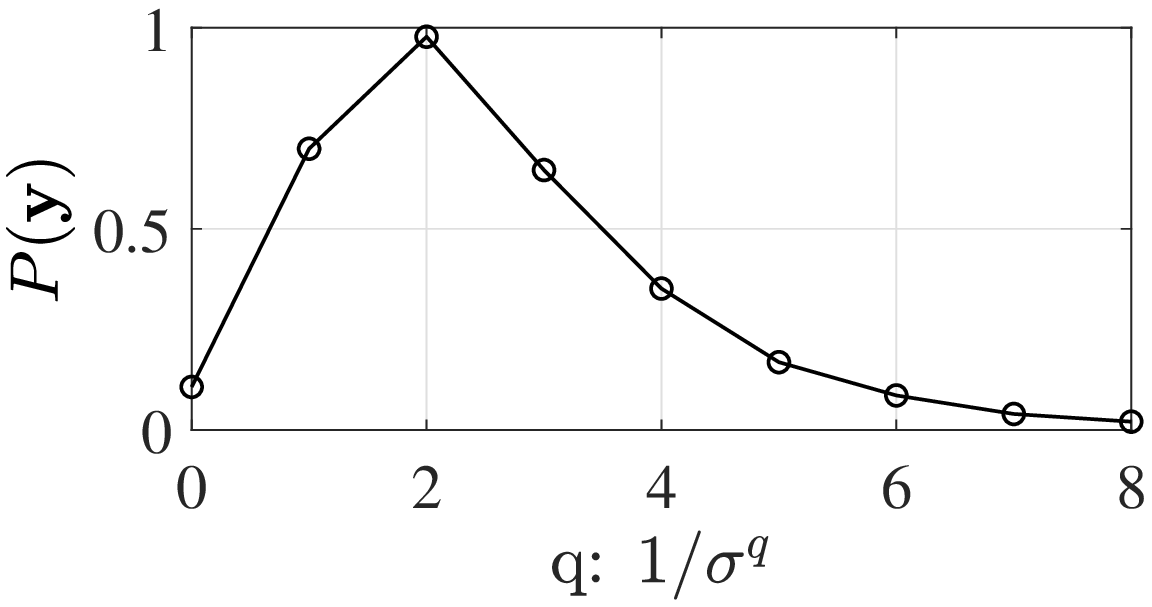}\end{minipage}  
& \hspace{0.0cm}\begin{minipage}{.275\textwidth}\vspace{0.5cm}\includegraphics[width=\linewidth]{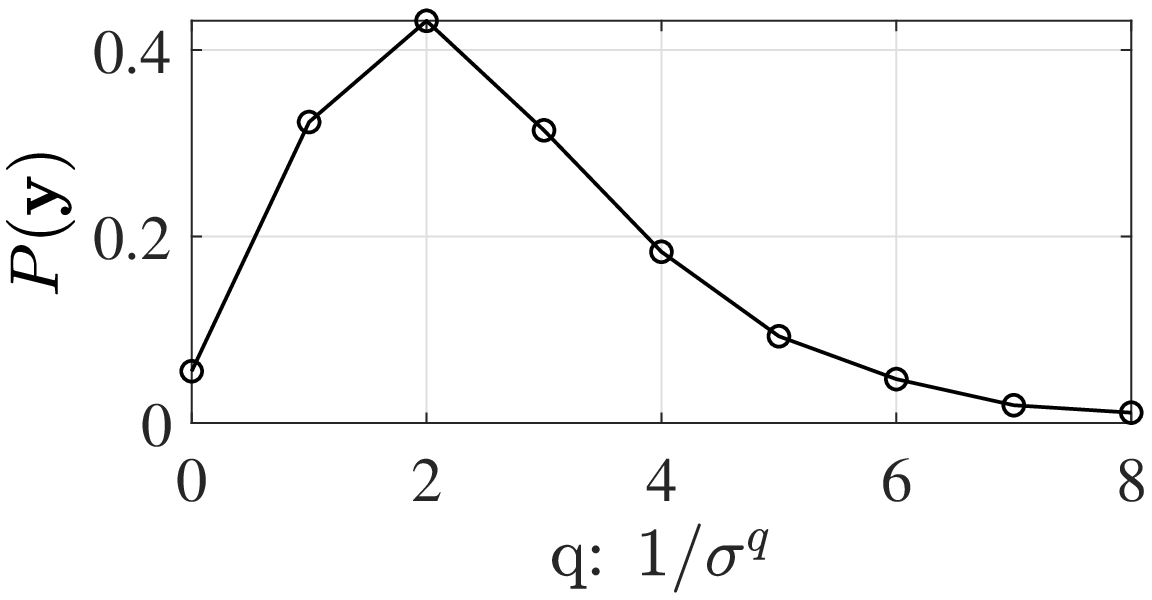}\end{minipage} 
& \hspace{0.0cm}\begin{minipage}{.275\textwidth}\vspace{0.5cm}\includegraphics[width=\linewidth]{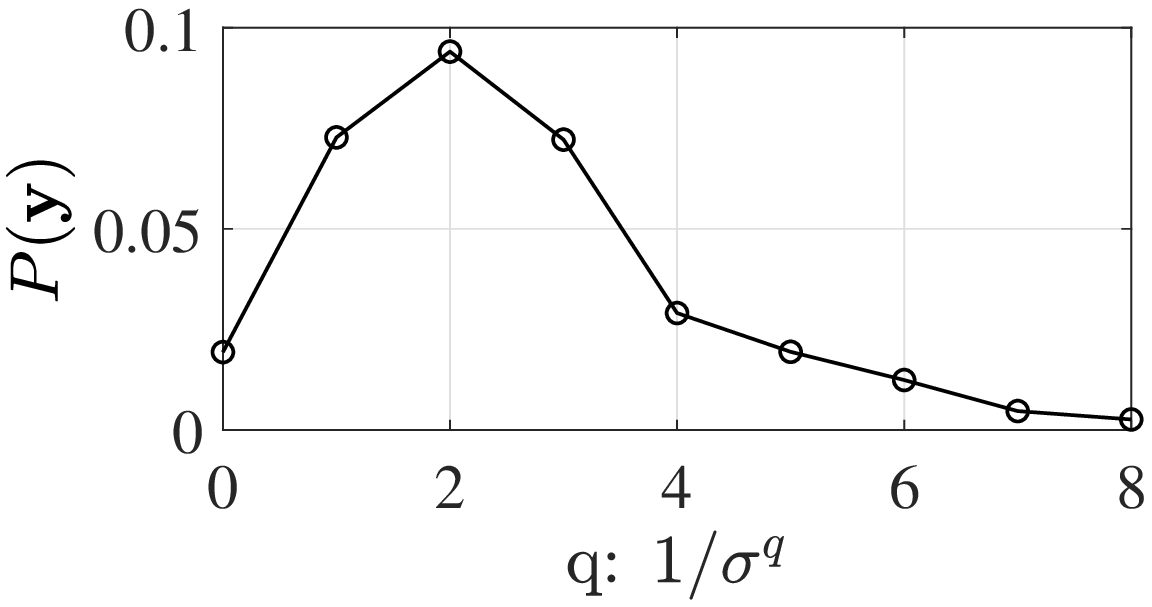}\end{minipage} \\
1 
& \hspace{0.0cm}\begin{minipage}{.275\textwidth}\vspace{0.5cm}\includegraphics[width=\linewidth]{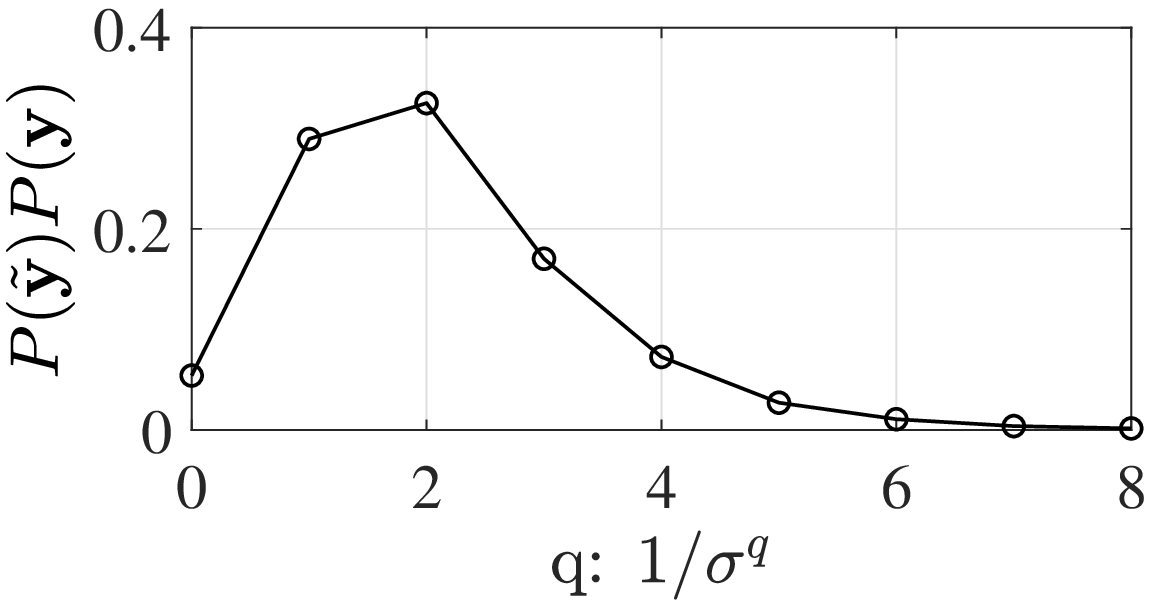}\end{minipage}  
& \hspace{0.0cm}\begin{minipage}{.275\textwidth}\vspace{0.5cm}\includegraphics[width=\linewidth]{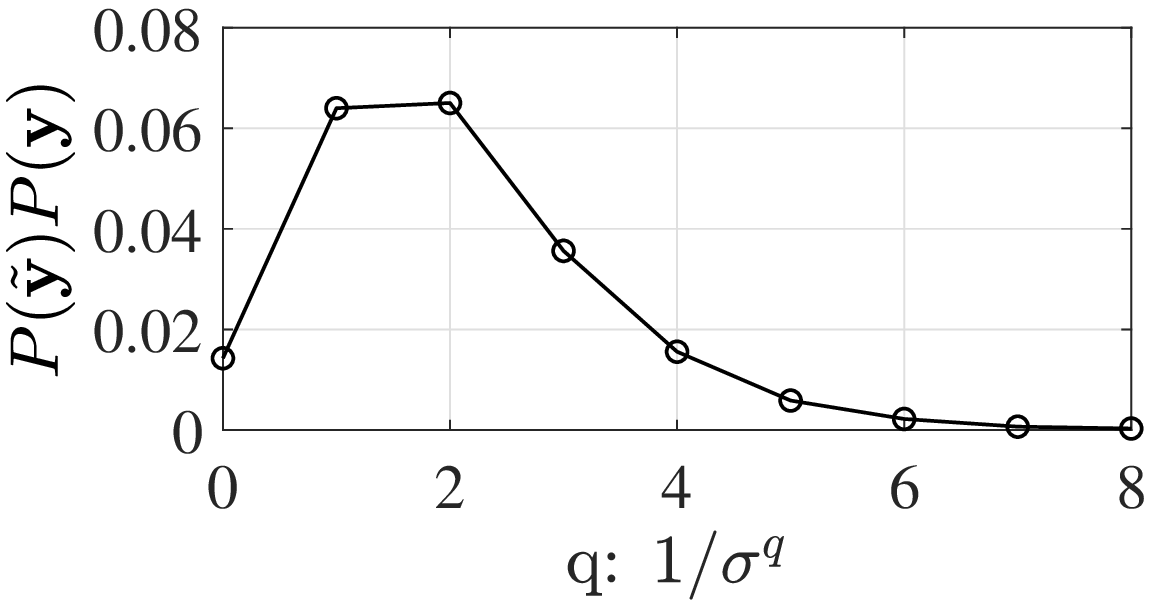}\end{minipage} 
& \hspace{0.0cm}\begin{minipage}{.275\textwidth}\vspace{0.5cm}\includegraphics[width=\linewidth]{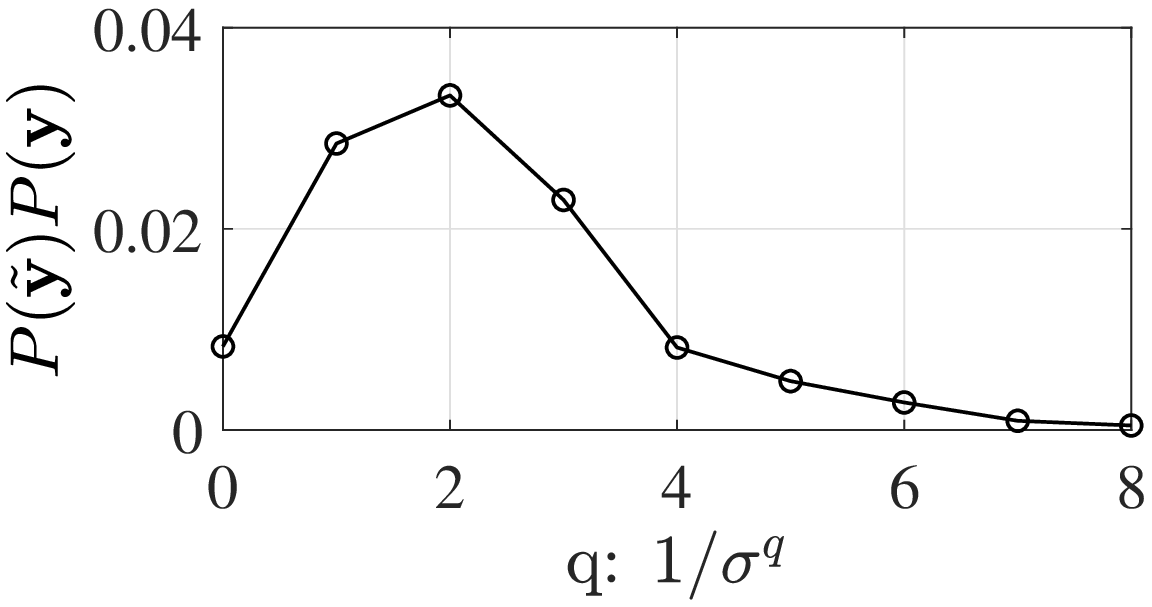}\end{minipage} \\
2 
& \hspace{0.0cm}\begin{minipage}{.275\textwidth}\vspace{0.5cm}\includegraphics[width=\linewidth]{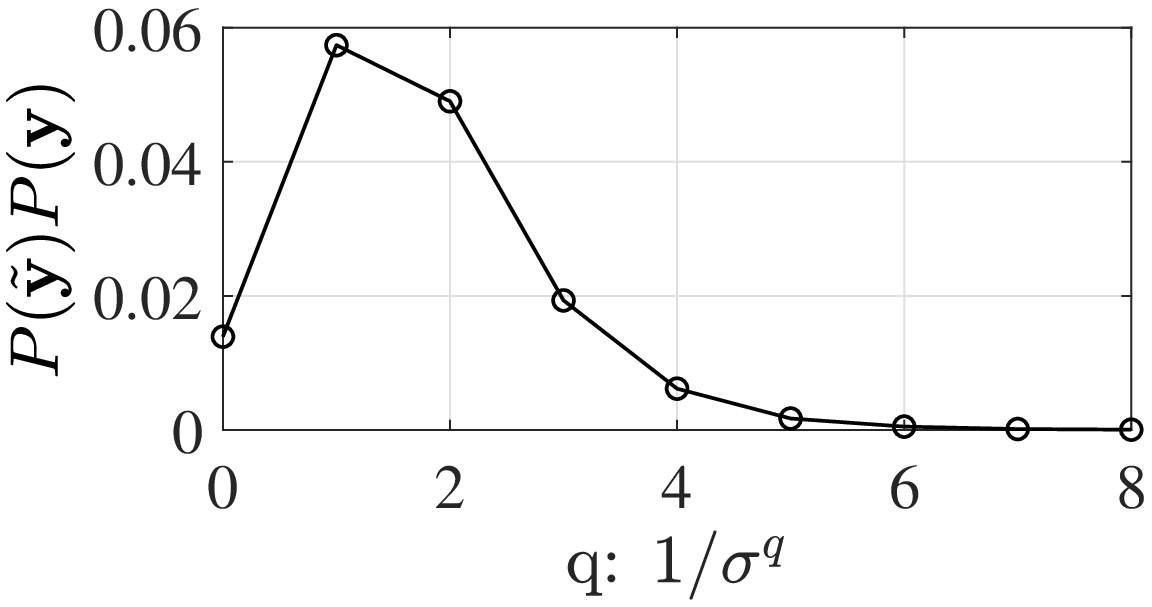}\end{minipage}  
& \hspace{0.0cm}\begin{minipage}{.275\textwidth}\vspace{0.5cm}\includegraphics[width=\linewidth]{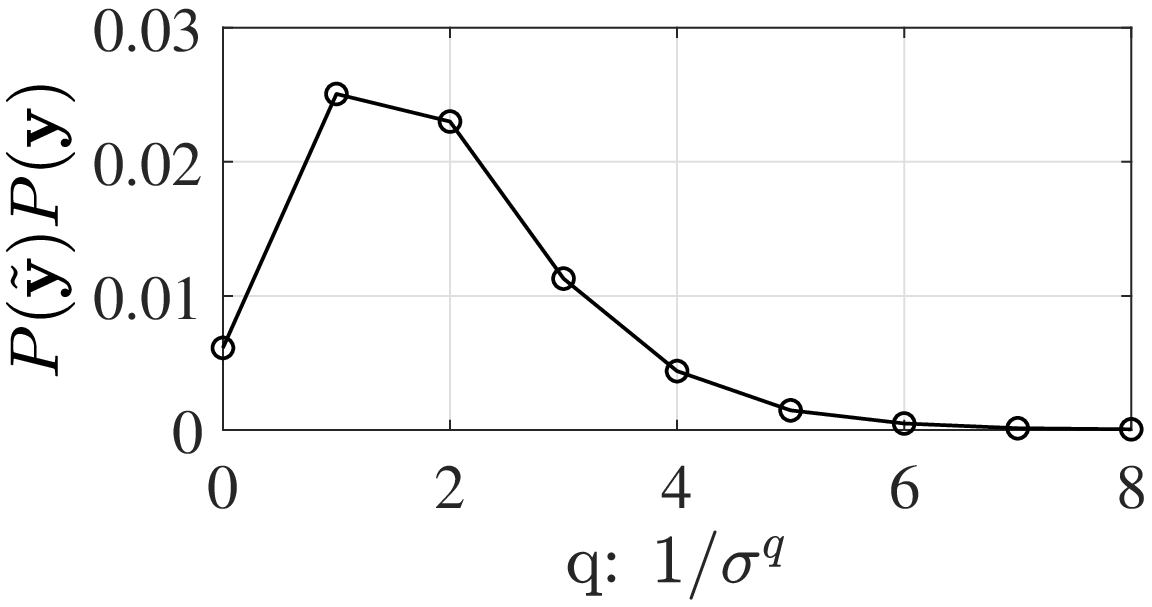}\end{minipage} 
& \hspace{0.0cm}\begin{minipage}{.275\textwidth}\vspace{0.5cm}\includegraphics[width=\linewidth]{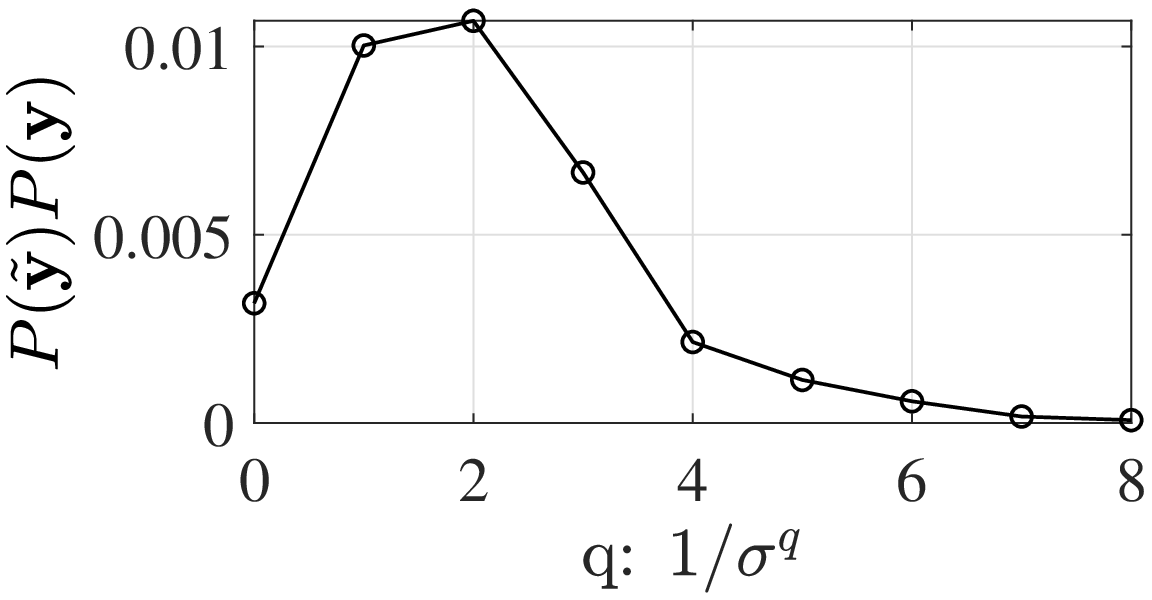}\end{minipage} \\
3 
& \hspace{0.0cm}\begin{minipage}{.275\textwidth}\vspace{0.5cm}\includegraphics[width=\linewidth]{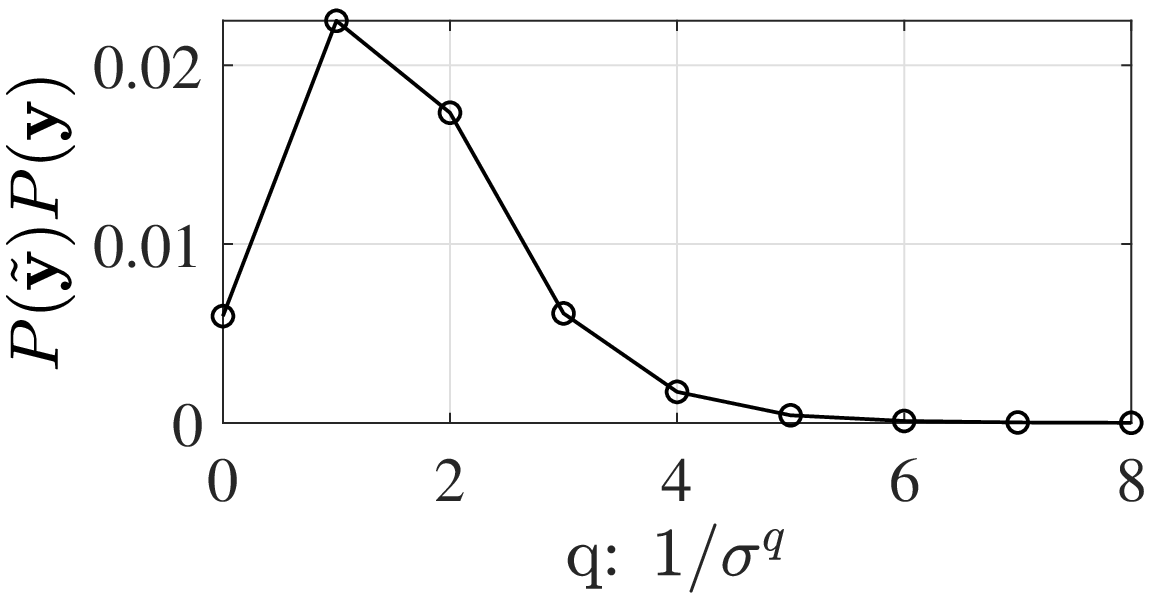}\end{minipage}  
& \hspace{0.0cm}\begin{minipage}{.275\textwidth}\vspace{0.5cm}\includegraphics[width=\linewidth]{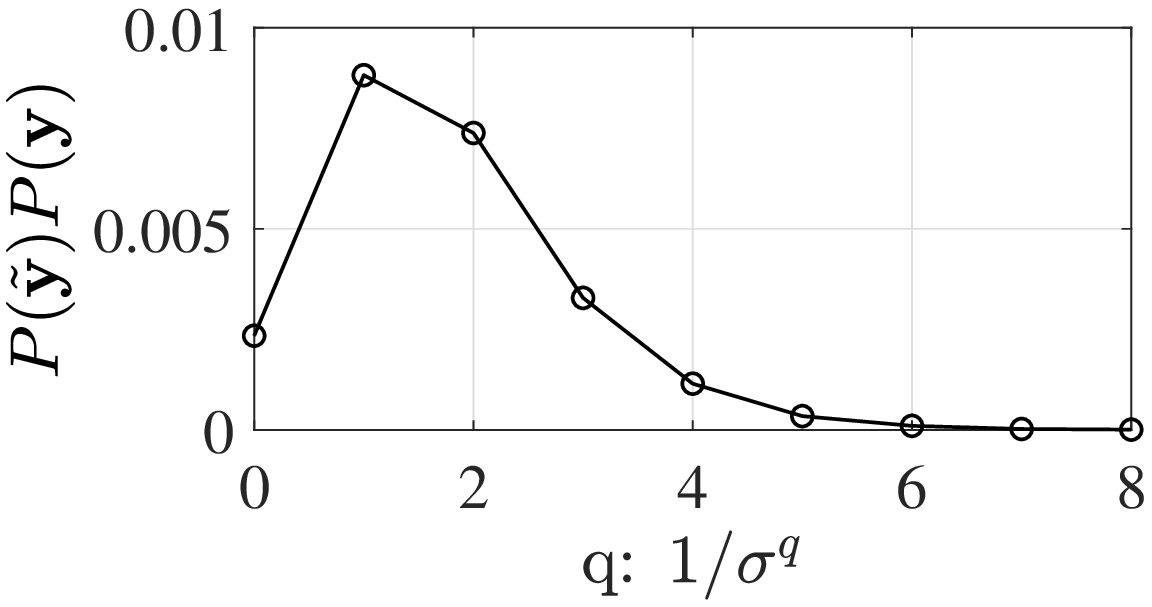}\end{minipage} 
& \hspace{0.0cm}\begin{minipage}{.275\textwidth}\vspace{0.5cm}\includegraphics[width=\linewidth]{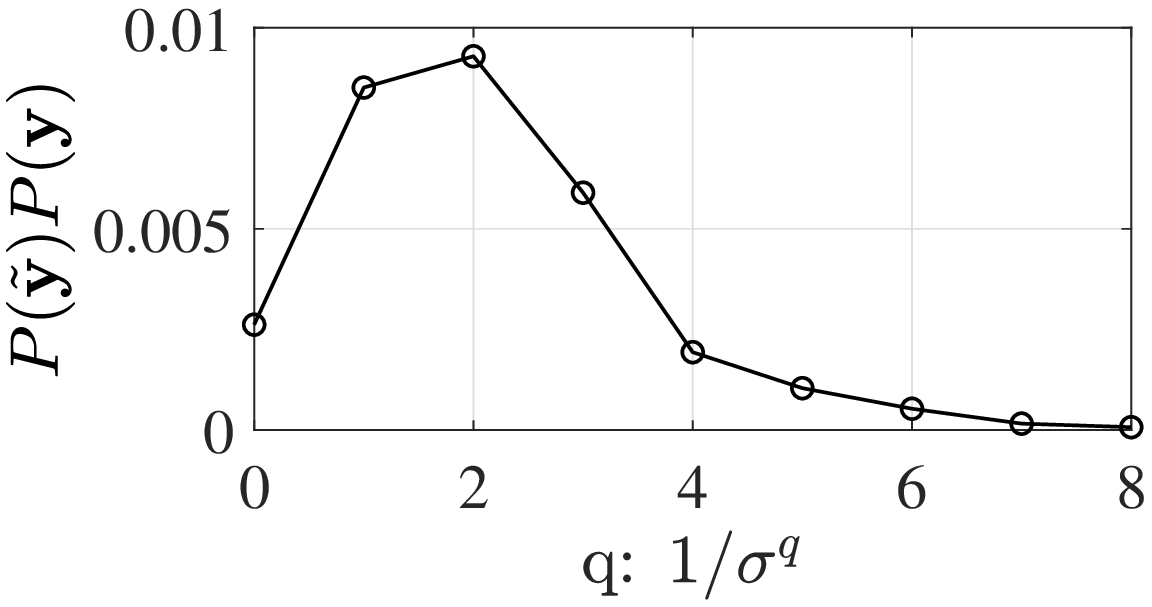}\end{minipage} \\
4 
& \hspace{0.0cm}\begin{minipage}{.275\textwidth}\vspace{0.5cm}\includegraphics[width=\linewidth]{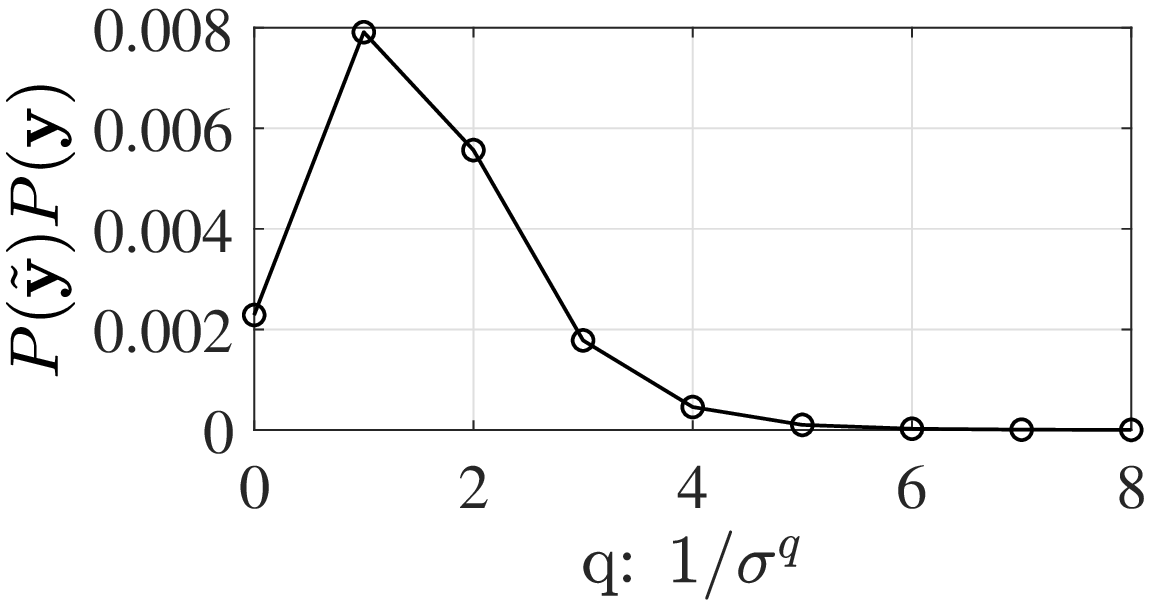}\end{minipage}  
& \hspace{0.0cm}\begin{minipage}{.275\textwidth}\vspace{0.5cm}\includegraphics[width=\linewidth]{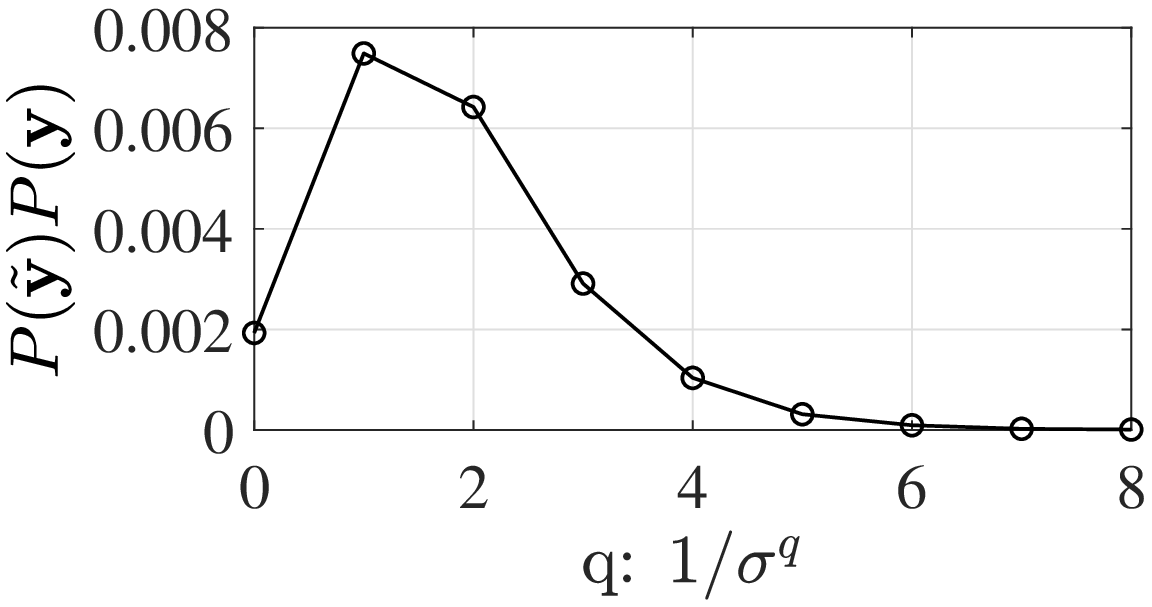}\end{minipage} 
& \hspace{0.0cm}\begin{minipage}{.275\textwidth}\vspace{0.5cm}\includegraphics[width=\linewidth]{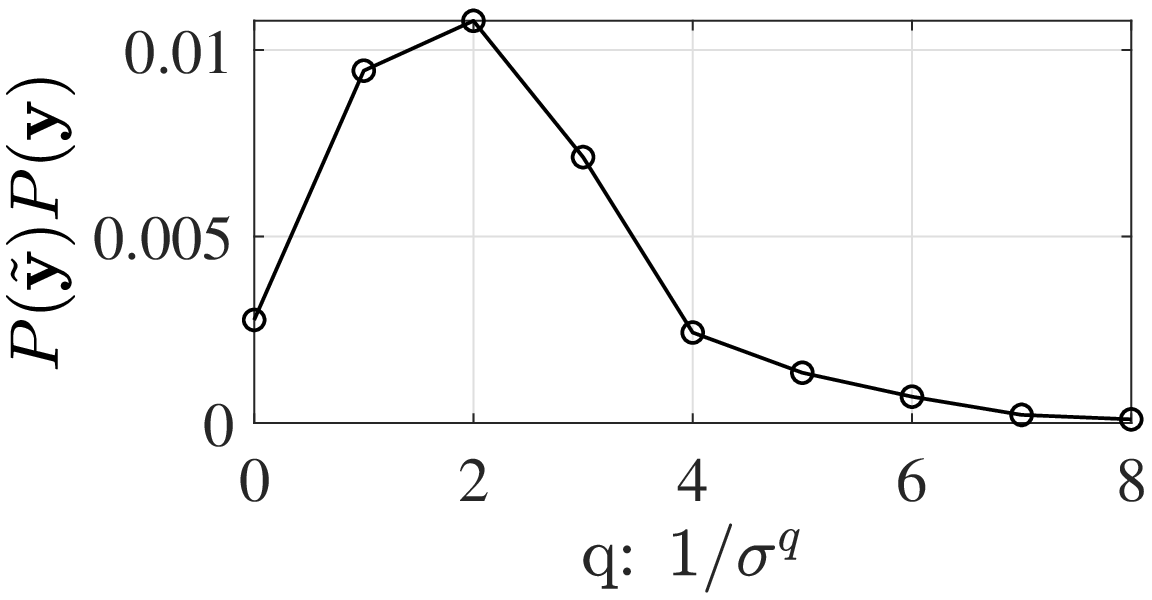}\end{minipage} \\
5 
& \hspace{0.0cm}\begin{minipage}{.275\textwidth}\vspace{0.5cm}\includegraphics[width=\linewidth]{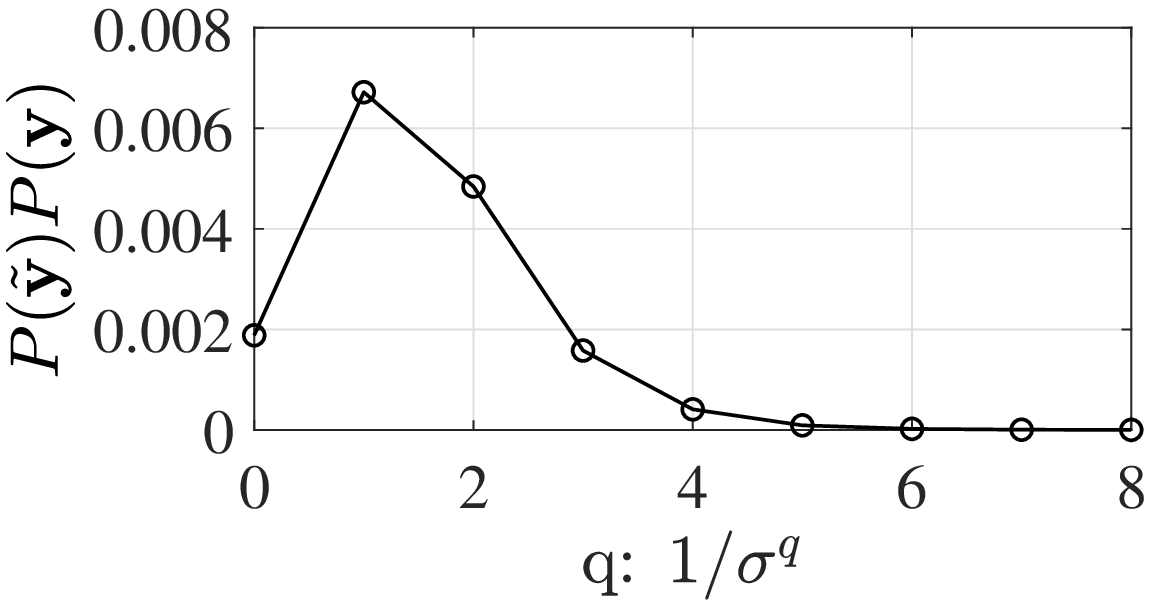}\end{minipage}  
& \hspace{0.0cm}\begin{minipage}{.275\textwidth}\vspace{0.5cm}\includegraphics[width=\linewidth]{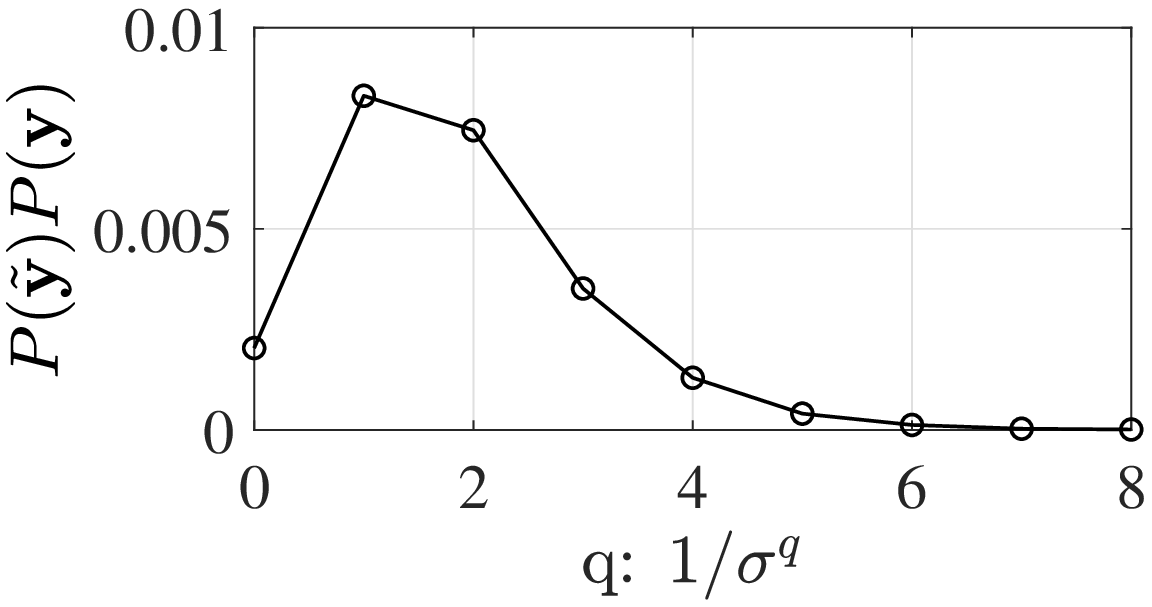}\end{minipage} 
& \hspace{0.0cm}\begin{minipage}{.275\textwidth}\vspace{0.5cm}\includegraphics[width=\linewidth]{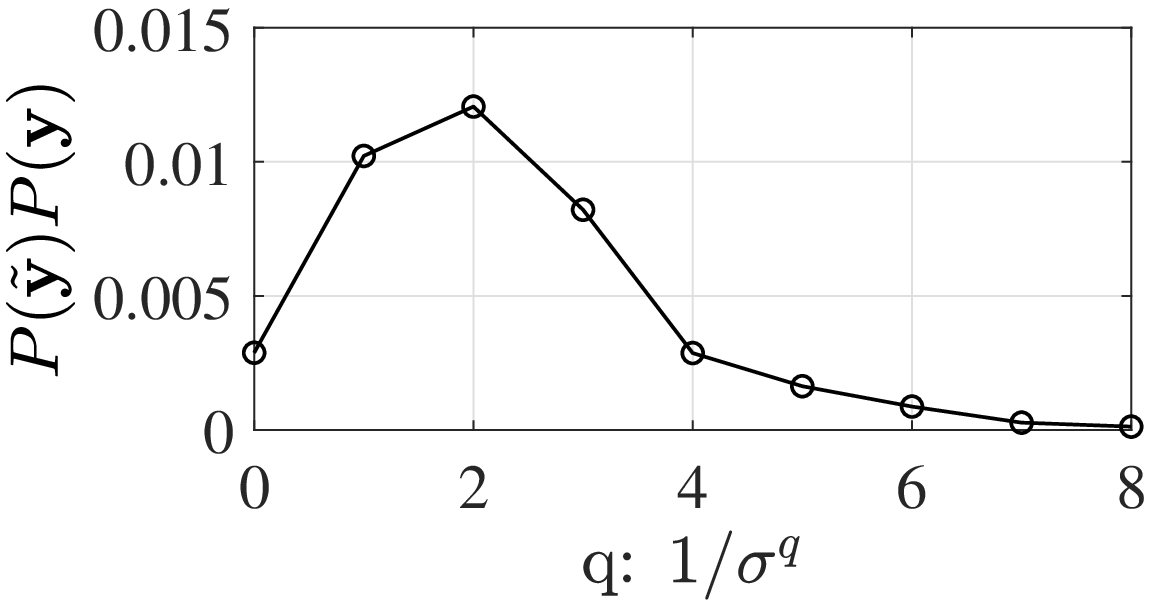}\end{minipage} \\
\vspace{0.0cm}
\end{tabular}
\label{tab:prior_predictive_performance}
\end{table*}

\subsection{A fatigue reliability problem}
The above numerical example reveals the advantage of using noninformative prior over the flat prior in probabilistic inference under small sample size. To examine its usefulness in realistic problems and further compare the performance of different noninformative priors, a fatigue life prediction problem is presented. 

The low cycle fatigue testing data reported in ASTM E739-10 \cite{ASTM2015E739} are used and shown in Table \ref{tab:sndata}. The cyclic strain amplitudes $(\Delta\varepsilon/2)$ are ($\sim$0.016, $\sim$0.0068, $\sim$0.0016, $\sim$0.0005), and each level has two or three data points. The following log-linear model is adopted,
\begin{equation}
\ln N = a_0 + a_1\ln\left(\Delta\varepsilon/2\right),
\label{eq:sn}
\end{equation}
where $a_0$ and $a_1$ are model parameters need to be identified. It should be noted that other variants of the above equation can be used to include material plasticity and mean stress effect etc. 
The discrepancy between the experimental data and the model prediction can be modeled using an uncertain variable with zero mean and a standard deviation of $\sigma_{e}$. To compare with regular least square method, a Gaussian likelihood is assumed and the Bayesian posterior is
\begin{equation}
\resizebox{\columnwidth}{!}{$
\begin{array}{rl}
p(a_0,a_1,\sigma_{e}|{\bm N}) & \propto \dfrac{1}{\sigma^{q}}\cdot\left(\dfrac{1}{\sigma^2} \right)^{n/2}\\
&\times \exp\left[ -\dfrac{\sum_{i=1}^{n}\left(\ln N_{i}-a_0-a_1\ln\left(\Delta\varepsilon_{i}/2\right) \right)^{2}}{2\sigma^2}\right].
\end{array}
$}
\label{eq:snpost}
\end{equation}
The exponent $q$ takes the values of $0,1,2,...$ to generate a flat prior and different noninformative priors, and $(\Delta\varepsilon_{i},N_{i})$, $i=1,...,n$ is the $i$th experimental data points of the total $n$ data points used for parameter estimation. 

\begin{figure*}[!h]
\centering
\subfloat[]{\label{fig:prior_fatigue_n8sigexp_neg2_2}\includegraphics[width=0.4\textwidth]{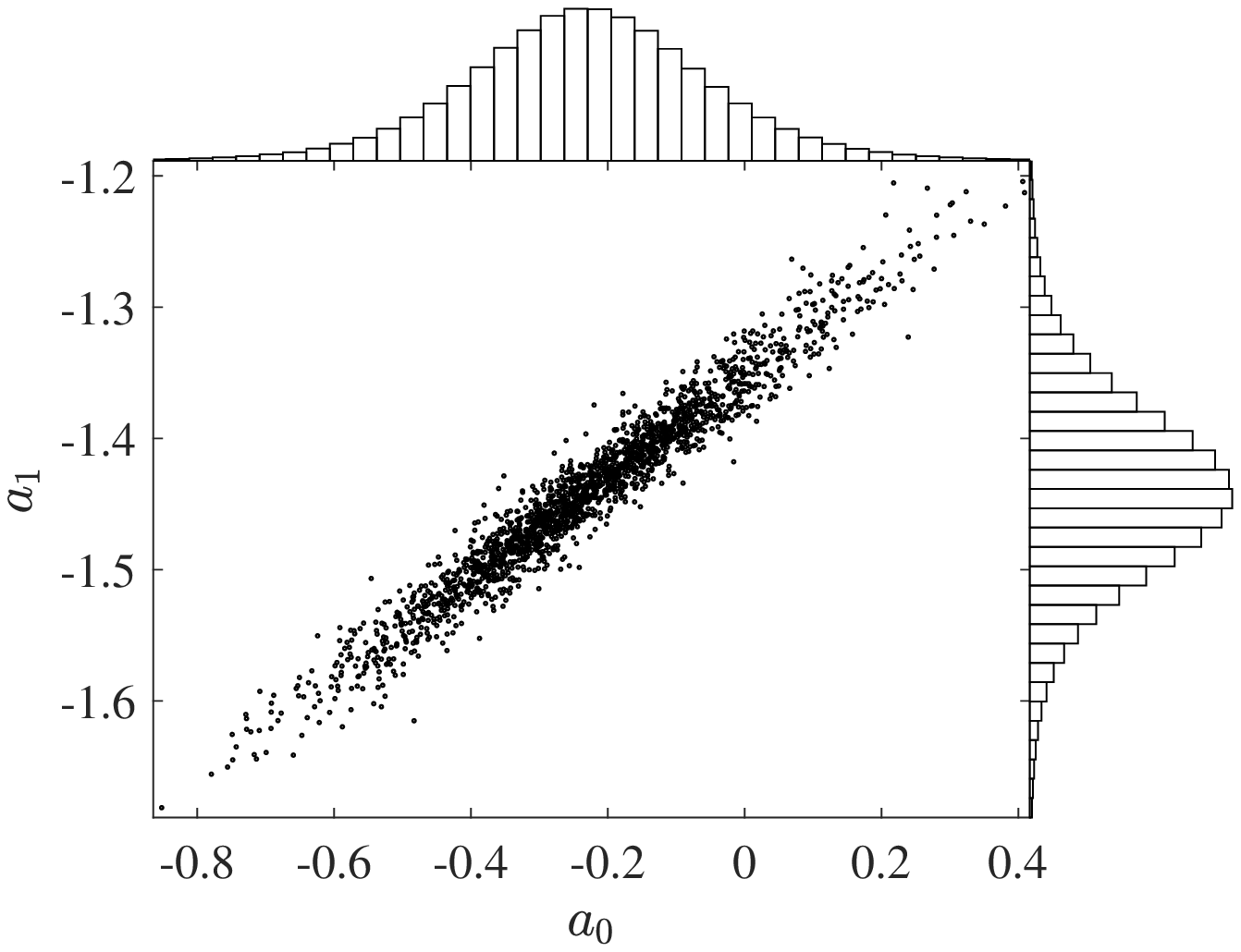}} \quad
\subfloat[]{\label{fig:prior_fatigue_n8sigexp_neg2_3}\includegraphics[width=0.4\textwidth]{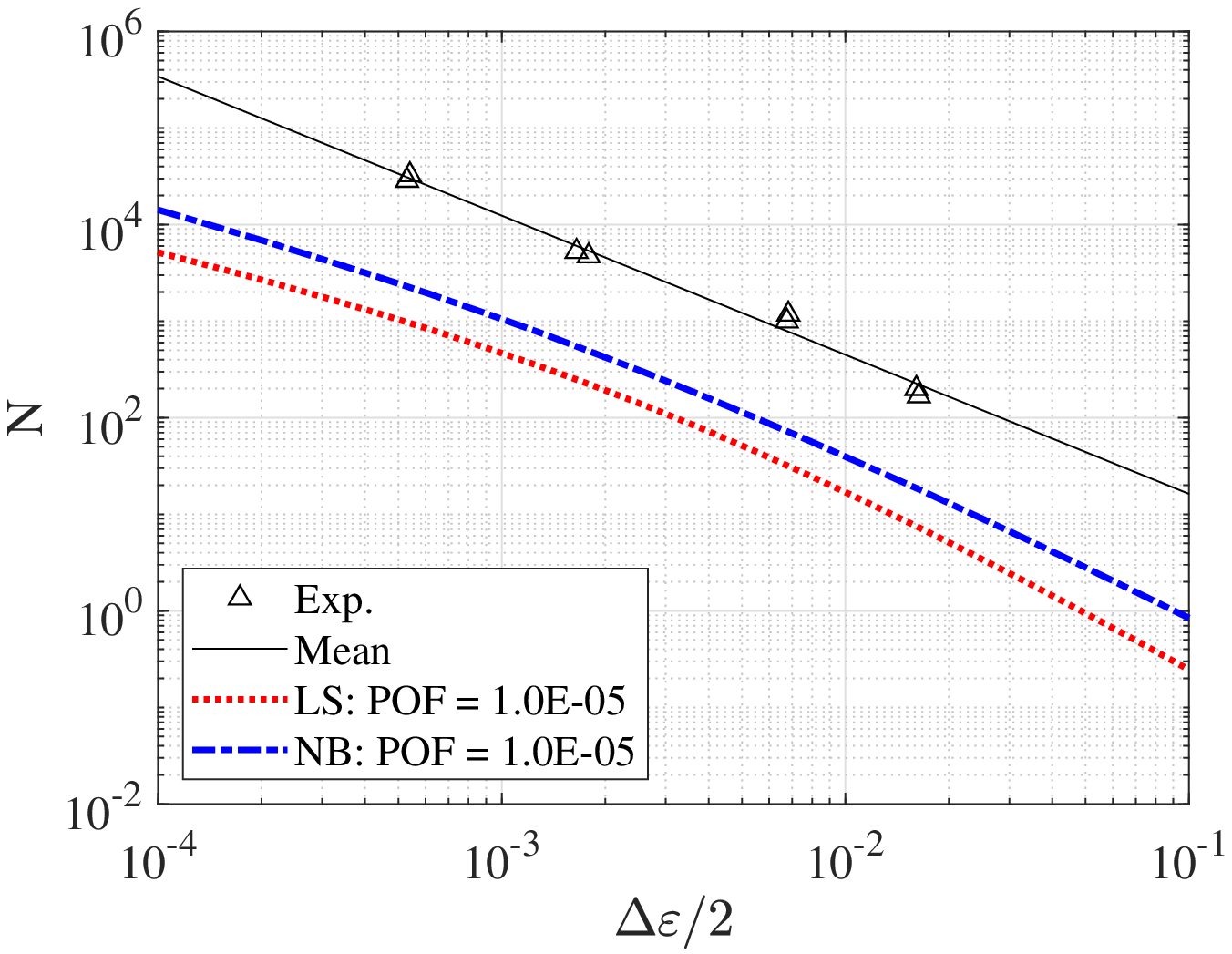}} \\
\caption{(a) MCMC samples drawn from noninformative Bayesian posterior with Jeffreys prior ($q=2$), and (b) the fatigue life results at $\mbox{POF}=10^{-5}$ obtained with regular least square (LS) estimator and noninformative Bayesian (NB).}%
\label{fig:plot_fatigue_performance}%
\end{figure*}

Notice that when $q=0$ the above equation reduces to a regular least square format. One data point ($i= 6$) is arbitrarily chosen for prediction performance evaluation and the rest eight data points are used to estimate the model parameters using Eq. (\ref{eq:snpost}).
\begin{table*}[!h]
\centering
\caption{Low cycle fatigue testing data. Source: Ref. \cite{ASTM2015E739}.}
\begin{tabular}{r r r r r r r r r r}
$i$ & 1 & 2 & 3 & 4 & 5 & 6 & 7 & 8 & 9 \\ \hline
$\Delta\varepsilon/2$ &0.01636 & 0.01609 & 0.00675 & 0.00682 & 0.00179 & 0.00160 & 0.00165 & 0.00053 & 0.00054 \\
$N$ & 168 & 200 & 1000 & 1180 & 4730 & 8035 & 5254 & 28617 & 32650 \\ \hline
\end{tabular}
\label{tab:sndata}
\end{table*}
To compare the performance of $1/\sigma^q$ type priors, $q=0,...5$ are used to represent regular least square, Jeffreys', and ALI priors. The global likelihood of fitting performance and prediction performance are evaluated using the method of Laplace approximation, and results are shown in Fig. \ref{fig:plot_fatigue_performance}. The global likelihood of fitting and prediction results indicate that the Jeffreys prior $1/\sigma^2$ outperforms the others. The result associated with $q=0$ is the Bayesian version of the regular least square estimation. 

\begin{figure*}[!h]
\centering
\subfloat[]{\label{fig:plot_fatigue_performance_n8_1}\includegraphics[width=0.4\textwidth]{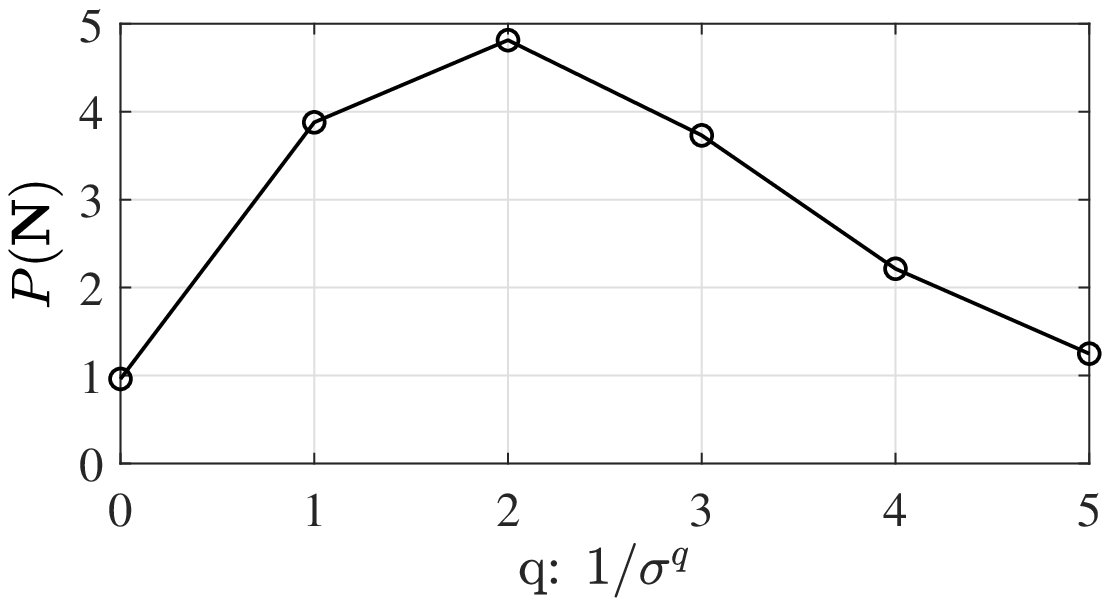}} \quad
\subfloat[]{\label{fig:plot_fatigue_performance_n8p1_2}\includegraphics[width=0.4\textwidth]{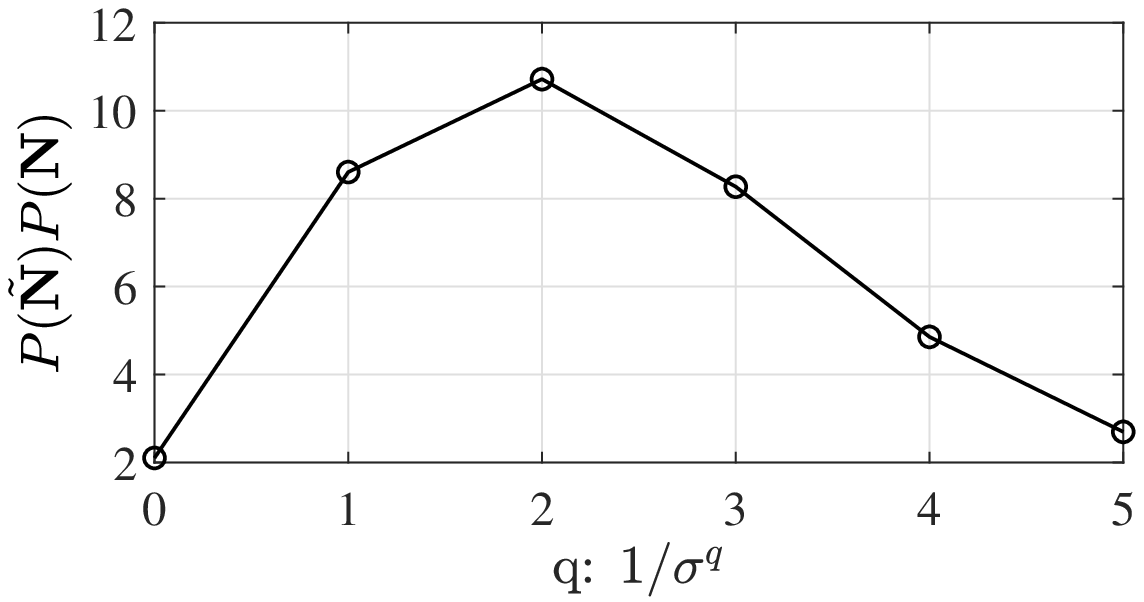}} \\
\caption{(a) The fitting performance $P({\bm N})$ evaluated with different $1/\sigma^{q}$ priors, and (b) the prediction performance $P(\tilde{{\bm N}})P({\bm N})$ evaluated with different $1/\sigma^{q}$ priors. One data point $\tilde{{\bm N}}$ is considered for prediction.}%
\label{fig:plot_fatigue_n8sigexp_neg2}%
\end{figure*}

To further demonstrate the difference between the regular least square and noninformative Bayesian approach under this sample size, the following reliability problem is considered. The fatigue life given a prescribed probability of failure (POF) is calculated for the cyclic strain range ($\Delta\varepsilon/2$) between $10^{-4}$ and $10^{-1}$. Given a cyclic strain range and a POF, the fatigue life $N_{\mbox{POF}}$ can be expressed as 
\begin{equation}
\mbox{Pr}(N>N_{\mbox{POF}}) = 1-\mbox{POF}.
\label{eq:pof}
\end{equation}
For regular least square estimator, it can be obtained using t-statistics given by Eq. (\ref{eq:LSbounds}) where $\gamma/2$ should be replaced with the prescribed POF because only the lower one-side bound is needed. For Bayesian with Jeffreys prior, the one-side bound can trivially be obtained using POF-quantile of the prediction results evaluated using MCMC samples of the posterior POF. Figure \ref{fig:plot_fatigue_n8sigexp_neg2} presents the comparisons of fatigue life results where the obvious difference is observed. The results of noninformative Bayesian is larger than those of regular least square, indicating a longer fatigue life for the same risk level, i.e., $10^{-5}$ POF. 

\section{Conclusion}
The study developed a noninformative Bayesian inference method for small sample problems. In particular, $1/\sigma^{q}$ type of priors were proposed to formulate the Bayesian posterior. It is shown that this type of prior represents the classical Jeffreys' prior (Fisher information), flat prior, and asymptotic locally invariant prior. More importantly, this type of priors were derived as the limiting states of Normal--Inverse--Gamma (NIG) conjugate priors, allowing for fast and efficient evaluation of the Bayesian posteriors and predictors in close--form expressions. To compare the performance of noninformative priors, a numerical linear regression problem and a realistic fatigue reliability problem were discussed in detail. The fitting and prediction performance measures in terms of global likelihood are used to evaluate different priors. It is observed that Jeffreys' prior, $1/\sigma^2$, yields the best fitting and prediction performance. Based on the current study, the following conclusions are drawn.
\begin{itemize}
	\item The $1/\sigma^{q}$ type of priors can be used as a reduced form of NIG conjugate priors. The great features of conjugate priors, such as having analytical posterior and prediction PDFs, can be retained when using noninformative priors for Bayesian linear regression analysis. The equivalence of NIG conjugate priors and $1/\sigma^{q}$ type of priors at the limiting state is firstly utilized in this study.
	\item For $1/\sigma^{q}$ type of noninformative priors, the classical Jeffreys prior $1/\sigma^{2}$ yields optimal fitting and prediction performance in terms of global likelihood or Bayes factors. When $q=0$ the Bayesian estimator can be viewed as the regular least square estimator. Results based on two case studies demonstrate the advantage of using noninformative Bayesian estimator with the $1/\sigma^{2}$ prior over the regular least square estimator under small sample size.
\end{itemize}

\section*{Acknowledgements}
The study was supported by the 
National Natural Science Foundation of China, Nos. 11872088, 51975546, U1930403.
 The support is gratefully acknowledged. The authors would like to thank the anonymous reviewers for their constructive comments.

\section*{Appendix}      
\numberwithin{equation}{subsection}
\label{sec:appendix}
\renewcommand{\thesubsection}{\Alph{subsection}}

\subsection{Conjugate prior, posterior, and prediction}\label{apdx:conjugate}
Assume a general linear model with a parameter vector $\bm{\theta}$ and a Gaussian likelihood with a scale parameter $\sigma^2$, i.e., the modeling error $\epsilon_{i} = \left(y_{i}-\mathbf{x}_{i}\bm{\theta}\right) \sim \mathrm{Norm}(0,\sigma^2)$. 
\subsubsection{Conjugate prior}
It is known that when the likelihood function belongs to the exponential family, a conjugate prior exists and belongs also to the exponential family. To obtain the joint conjugate using $p(\bm{\theta},\sigma^2)  = p(\bm{\theta}|\sigma^2)p(\sigma^2)$, it is necessary to look at the conjugate prior of $\sigma^2$ first. For a Gaussian likelihood with an unknown variance $\sigma^2$ and zero mean, the conjugate prior for $\sigma^2$ is an inverse Gamma distribution, i.e., $p(\sigma^{2}) \sim \mathrm{IG}(\alpha,\beta)$, and is expressed as
\begin{equation}
p(\sigma^2) = \frac{\beta^{\alpha}}{\Gamma(\alpha)}\left(\frac{1}{\sigma^2}\right)^{\alpha+1} \cdot \exp\left(-\frac{\beta}{\sigma^2} \right),
\label{eq:ig}
\end{equation}
where $\alpha>0$ and $\beta>0$ are shape and scale parameters, respectively. The posterior distribution of $\sigma^2$, given $n$ data points, has the following new shape and scale parameters
\begin{equation}
\begin{array}{rl}
(\alpha',\beta') &= \displaystyle{\left(\alpha + \frac{n}{2},\beta+\frac{\sum_{i}^{n}\left( y_{i}-\mathbf{x}_{i}\bm{\theta}\right)^2}{2} \right)}\\
& = \displaystyle{\left(\alpha+\frac{n}{2},\beta + \frac{\mathrm{SSE}}{2} \right)},
\end{array}
\label{eq:newpar}
\end{equation}
where $\mathrm{SSE}=\sum_{i}^{n}\left( y_{i}-\mathbf{x}_{i}\bm{\theta}\right)^2$ is the sum of squared errors. 
Consider the joint conjugate of (${\bm\theta},\sigma^2$), a widely-adopted conjugate priors for Bayesian linear regression is the Normal-Inverse-Gamma (NIG) prior for $(\bm{\theta},\sigma^2)$, and can be written as Eq. (\ref{eq:nig}).
\begin{equation}
\resizebox{\columnwidth}{!}{$
\begin{aligned}
p(\bm{\theta},\sigma^2) & = p(\bm{\theta}|\sigma^2)p(\sigma^2) = 
\mathrm{NIG}(\bm{\mu_{\theta}},\bm{\Sigma_{\theta}},\alpha,\beta) \\
& = Z\cdot\exp\left\{-\frac{1}{\sigma^2}\left[\beta+\frac{1}{2}\left(\bm{\theta}-\bm{\mu_{\theta}} \right)^{T}\bm{\Sigma_{\theta}}^{-1}\left(\bm{\theta}-\bm{\mu_{\theta}} \right) \right] \right\} \\
& \propto Z' \cdot\exp\left\{-\frac{1}{\sigma^2}\left[\beta+\frac{1}{2}\left(\bm{\theta}-\bm{\mu_{\theta}} \right)^{T}\bm{\Sigma_{\theta}}^{-1}\left(\bm{\theta}-\bm{\mu_{\theta}} \right) \right] \right\},
\end{aligned}
$}
\label{eq:nig}
\end{equation}
where 
\begin{equation}
Z=\frac{\beta^{\alpha}}{(2\pi)^{k/2}\Gamma(\alpha)\sqrt{\left|\bm{\Sigma_{\theta}} \right|}}\left( \frac{1}{\sigma^2}\right)^{\alpha+k/2+1},
\label{eq:Z}
\end{equation}
and 
\begin{equation}
Z'=\left(\frac{1}{\sigma^2} \right)^{\alpha+k/2+1}.
\label{eq:Zprime}
\end{equation}

Using the inverse gamma distribution for $\sigma^{2}$, the initial distribution parameter $(\alpha,\beta)$ still needs to be determined. Unfortunately there is no formal rules to do that. \citet{gelman2006prior} discussed the inverse gamma distribution $\mathrm{IG}(\epsilon,\epsilon)$ as an attempt at noninformativeness within the conjugate family, with $\epsilon$ set to a low value such as 1 or 0.01 or 0.001. A difficulty of this prior is that $\epsilon$ must be set to a reasonable value. Inferences become very sensitive to $\epsilon$ for cases where low values of $\sigma^2$ are possible. Due to this reason, the $\mathrm{IG}(\epsilon,\epsilon)$ family of prior for $\sigma^2$ is not recommended by the author. It is further argued that the resulting inference basing on it for cases where $\sigma^2$ is estimated near zero are the cases that classical and Bayesian inferences differ the most \cite{gelman2006prior}. \citet{browne2006comparison} also mentioned this disadvantage and suggested using a uniform one $\mathrm{U}(0,1/\epsilon)$ to reduce this defect. However, choosing an appropriate value for $\epsilon$ for $\mathrm{U}(0, 1/\epsilon)$ still requires attentions and understanding of the data. Indeed the effect of prior and its parameters are highly subjected to the data and problem, and its sensitivity to Bayesian inferences based on sparse data or extensive data can be dramatically different as the data dominants the posterior due to {\it The Law of Large Numbers}. 

\subsubsection{Posterior distribution}
The posterior distribution of $(\bm{\theta},\sigma^{2})$ is expressed as
\begin{equation}
p(\bm{\theta},\sigma^2|\mathbf{y}) = \dfrac{p(\bm{\theta},\sigma^2)p(\mathbf{y}|\bm{\theta},\sigma^2)}{p(\mathbf{y})}.
\label{eq:postpdf}
\end{equation}
The term $p(\mathbf{y})=\int p(\bm{\theta},\sigma^2)p(\mathbf{y}|\bm{\theta},\sigma^2)\mathrm{d}\bm{\theta}\mathrm{d}\sigma^2$ is the marginal distribution of the data. It can be shown that after some algebraic operations, the posterior can be written as
\begin{equation}
\resizebox{1.0\columnwidth}{!}{$
\begin{array}{rl}
 p(\bm{\theta},\sigma^2|\mathbf{y}) &\propto \displaystyle\left(\frac{1}{\sigma^2}\right)^{\alpha^{*}+k/2+1}\\
&\times \displaystyle\exp\left\{-\frac{1}{\sigma^2}\left[\beta^{*} + \frac{1}{2}\left(\bm{\theta}-\bm{\mu}^{*}\right)^{T}\bm{\Sigma}^{*-1}\left(\bm{\theta}-\bm{\mu}^{*}\right) \right] \right\},
\end{array}
$}
\label{eq:postprop}
\end{equation}
where
\begin{equation*}
\left\{
\begin{array}{rl}
\bm{\mu}^{*}&=\left(\bm{\Sigma}^{-1}+\mathbf{x}^{T}\mathbf{x} \right)^{-1}\left(\bm{\Sigma}^{-1}\bm{\mu}+\mathbf{x}^{T}\mathbf{y} \right) \\
\bm{\Sigma}^{*} &= \left(\bm{\Sigma}^{-1}+\mathbf{x}^{T}\mathbf{x} \right)^{-1} \\
\alpha^{*}&=\alpha+n/2 \\
\beta^{*}&=\beta+\frac{1}{2}\left(\bm{\mu}^{T}\bm{\Sigma}^{-1}\bm{\mu}+\mathbf{y}^{T}\mathbf{y}-\bm{\mu}^{*T}\bm{\Sigma}^{*-1}\bm{\mu}^{*} \right)
\end{array}
\right. .
\label{eq:betastar}
\end{equation*}
The initial NIG conjugate parameter $({\bm\mu},{\bm\Sigma},\alpha,\beta)$ is updated to a new set of parameters $({\bm\mu}^{*},{\bm\Sigma}^{*},\alpha^{*},\beta^{*})$.

\subsubsection{Prediction}
The Bayesian prediction of the response given the data $\mathbf{y}$ can be obtained using above distributions with a new set of input variable $\tilde{\mathbf{x}}$. Denote the corresponding prediction as $\tilde{\mathbf{y}}$. The prediction posterior distribution of $\tilde{\mathbf{y}}$ can be expressed as
\begin{equation}
\begin{aligned}
p(\tilde{\mathbf{y}}|\mathbf{y}) &= \int p(\tilde{\mathbf{y}}|\bm{\theta},\sigma^{2})p(\bm{\theta},\sigma^{2}|\mathbf{y})\mathrm{d}\bm{\theta}\mathrm{d}\sigma^2 \\
&= \int \mathrm{MVN}(\tilde{\mathbf{x}}\bm\theta,\sigma^{2}\mathbf{I})\times\mathrm{NIG}(\bm{\mu}^{*},\bm{\Sigma}^{*},\alpha^{*},\beta^{*})\mathrm{d}\bm{\theta}\mathrm{d}\sigma^{2} \\
&=\mathrm{MVT}_{2\alpha^{*}}\left(\tilde{\mathbf{x}}\bm{\mu}^{*},\frac{\beta^{*}}{\alpha^{*}}(\mathbf{I} + \tilde{\mathbf{x}}\bm{\Sigma}^{*}\tilde{\mathbf{x}}^{T}) \right).
\end{aligned}
\label{eq:prepost}
\end{equation}
When a single point is evaluated, Eq. (\ref{eq:prepost}) is reduced to a one-dimensional Student distribution, and the confidence interval is readily evaluated using the inverse Student distribution. For multiple point simultaneous evaluations, the interval contours can be approximated using a few methods as suggested in Ref. \cite{kotz2004multivariate}. 


To compare, the simple linear regression results of the prediction mean and bounds are given. The prediction mean is $\tilde{\mathbf{x}}\hat{\bm\theta}$, and the prediction bounds of level $(1-\gamma)\%$ are
\begin{equation}
\tilde{\mathbf{y}}_{1-\gamma} = \tilde{\mathbf{x}}\hat{\bm\theta}\; \pm \;\mathrm{tinv}(\gamma/2,n-k) \cdot\mathrm{\bm\sigma_{\mathbf{y}}},
\label{eq:LSbounds}
\end{equation}
where $\hat{\bm\theta}=(\mathbf{x}^{T}\mathbf{x})^{-1}(\mathbf{x}^{T}\mathbf{y})$ is the maximum likelihood estimator of $\bm\theta$. It is known that $\hat{\bm\theta}$ is also identical to the least square estimator and the Bayesian estimator, i.e., Eq. (\ref{eq:postthetatheory}). The term $\mathrm{tinv}(\gamma/2,n-k)$ represents the $\gamma/2$-percentile value of the standard Student's t-distribution with $n-k$ degrees-of-freedom, and $\bm\sigma_{\mathbf{y}}$ is the standard error of the prediction given by
\begin{equation}
\bm\sigma_{\mathbf{y}}^{2}= \frac{\mathbf{e}^{T}\mathbf{e}}{n-k} \cdot \left[\mathbf{I}+\tilde{\mathbf{x}}(\mathbf{x}^{T}\mathbf{x})^{-1}\tilde{\mathbf{x}}^{T} \right],
\label{eq:s2}
\end{equation}
where the term $\mathbf{e}=\mathbf{y}-\mathbf{x}\hat{\bm\theta}$ is the discrepancy vector between model and observation, also called residual or error. As multiple points are estimated, F-statistic can be used instead of t-statistics to yield the Working-Hotelling confidence intervals \cite{graybill1967linear},
\begin{equation}
\tilde{\mathbf{y}}_{1-\gamma} = \tilde{\mathbf{x}}\hat{\bm\theta}\; \pm \;\sqrt{2\mathrm{Finv}(\gamma,k,n-k)} \cdot\mathrm{\bm\sigma_{\mathbf{y}}},
\label{eq:wh}
\end{equation} 
where $\mbox{Finv}(\gamma,k,n-k)$ represents the $\gamma$-percentile value of the F-distribution with $(k,n-k)$ degrees-of-freedom. 

\subsection{Fisher information matrix and Kullback--Leibler divergence}\label{app:fishkl}
The equivalence of Fisher information matrix and Kullback--Leibler divergence can be shown as follows. The matrix form of Fisher information matrix of a continuous PDF with a parameter vector $\bm\theta$ writes, 
\begin{equation}
\mathbf{I}(\bm\theta) = \mathbb{E}_{\bm{\theta}}\left[\nabla \ln p_{\bm\theta}(x) \cdot \nabla \ln p_{\bm\theta}(x)^{T} \right].
\label{eq:fim}
\end{equation}
Under the condition that the integration and derivation may be exchanged and the log-function is twice differentiable, the following Eq. (\ref{eq:app_eq1}) can be attained.

\begin{equation}
\begin{array}{rl}
\nabla^2\ln p_{\bm\theta}(x) 
& = \displaystyle\frac{\nabla^2 p_{\bm\theta}(x)}{p_{\bm\theta}(x)}-\frac{\nabla p_{\bm\theta}(x)\nabla \cdot p_{\bm\theta}(x)^{T}}{\left[p_{\bm\theta}(x)\right]^2}  \\
& =\displaystyle\frac{\nabla^2 p_{\bm\theta}(x)}{p_{\bm\theta}}-\nabla\ln p_{\bm\theta}(x)\cdot \nabla\ln p_{\bm\theta}(x)^{T}.
\label{eq:app_eq1}
\end{array}
\end{equation}
It is noted that
\begin{equation}
\mathbb{E}_{\bm\theta}\left[\frac{\nabla^2 p_{\bm\theta}(x)}{p_{\bm\theta}(x)}\right] = \int\nabla^{2}p_{\bm\theta}(x)\mathrm{d}x = \nabla^{2}1 = 0.
\label{eq:app_eq3}
\end{equation}
The following result, that is, the negative expected Hessian matrix of log-function is equal to the Fisher information matrix, can be obtained.
\begin{equation}
\mathbf{I}(\bm\theta) = \mathbb{E}_{\bm\theta}\left[\nabla\ln p_{\bm\theta}(x)\cdot\nabla\ln p_{\bm\theta}(x)^{T}\right] 
=-\mathbb{E}_{\bm\theta}\left[\nabla^2\ln p_{\bm\theta}(x)\right],
\label{eq:app_eq2}
\end{equation}
where $\nabla^2\ln p_{\bm\theta}(x)$ is recognized as the Hessian matrix of $\ln p_{\bm\theta}(x)$, denoted as ${\mathbf{H}}_{\ln p_{\bm\theta}(x)}$. The Fisher information matrix can loosely be thought as the curvature matrix of the log-function graph.

The Kullback--Leibler divergence, also called the relative entropy, between two distributions can be written as
\begin{equation}
\mathbf{D}(\bm\theta||\bm{\theta'})=\int p_{\bm\theta}(x)\frac{\ln p_{\bm\theta}(x)}{\ln p_{\bm\theta'}(x)}\mathrm{d}x=\mathbb{E}_{\bm\theta}\left[\ln p_{\bm\theta}(x)-\ln p_{\bm\theta'}(x)\right].
\label{eq:kl}
\end{equation}
Expand $\ln p_{\bm\theta'}(x)$ around ${\bm\theta}$ to the second order to have,
\begin{equation}
\begin{array}{rl}
\ln p_{\bm\theta'}(x) & \approx \ln p_{\bm\theta}(x) + (\bm\theta-\bm\theta')\nabla\ln p_{\theta}(x) \\
& + \displaystyle\frac{1}{2}(\bm\theta-\bm\theta')^{T}\nabla^2\ln p_{\bm\theta}(x)(\bm\theta-\bm\theta')
\end{array}.
\label{eq:kl1}
\end{equation}
Substitute Eq. (\ref{eq:kl1}) into Eq. (\ref{eq:kl})
\begin{equation}
\resizebox{\columnwidth}{!}{$
\mathbf{D}(\bm\theta||\bm\theta') \approx -\mathbb{E}_{\bm\theta}\left[\Delta\bm\theta \cdot \nabla\ln p_{\bm\theta}(x)\right] - \displaystyle\frac{1}{2}\mathbb{E}_{\bm\theta}\left[\Delta\bm\theta\cdot \nabla^2\ln p_{\bm\theta}(x)\cdot \Delta\bm\theta^{T} \right],
$}
\label{eq:kl3}
\end{equation}
where $\Delta\bm\theta=(\bm\theta-\bm\theta')$.
The first term of the right hand side of Eq. (\ref{eq:kl3}) is zero and the second term of the right hand side is related to Eq. (\ref{eq:app_eq2}). The KL divergence can finally be expressed as,
\begin{equation}
\mathbf{D}(\bm\theta||\bm\theta') \approx \frac{1}{2}\Delta\bm\theta^{T}\cdot \mathbf{I}(\bm\theta) \cdot \Delta\bm\theta.
\label{eq:kl4}
\end{equation}

Furthermore, differentiation of Eq. (\ref{eq:kl}) with respect to $\bm\theta'$ to obtain
\begin{equation}
\begin{array}{rl}
\nabla_{{\bm\theta'}} \mathbf{D}({\bm\theta}||{\bm\theta'}) & = \nabla_{{\bm\theta'}} \mathbb{E}_{\bm\theta}\left[\ln p_{\bm\theta}(x) - \ln p_{\bm\theta'}(x) \right]\\
& = -\mathbb{E}_{\bm\theta}\left[\displaystyle\frac{\nabla p_{\bm\theta'}(x)}{p_{\bm\theta'}(x) }\right]
\end{array}.
\label{eq:kl5}
\end{equation}
Continue to differentiate Eq. (\ref{eq:kl5}) with respect to $\bm\theta'$ to obtain
\begin{equation}
\nabla_{{\bm\theta'}}^2 \mathbf{D}({\bm\theta}||{\bm\theta'}) = -\mathbb{E}_{\bm\theta}\left[\nabla p_{\bm\theta'}(x)\frac{\nabla p_{\bm\theta'}(x)}{p_{\bm\theta}^2} -\frac{\nabla^2 p_{\bm\theta'}(x)}{p_{\bm\theta'}(x)} \right].
\label{eq:kl6}
\end{equation}
Notice that the expectation of the second term in Eq. (\ref{eq:kl6}), when $\bm\theta' = \bm\theta$, reads,
\begin{equation}
\mathbb{E}_{\bm\theta}\left[ \frac{\nabla^2 p_{\bm\theta'}(x)}{p_{\bm\theta'}(x)}\right] = \nabla^2\int p_{\bm\theta}(x) = \nabla^2 1=0,
\label{eq:kl7}
\end{equation}
and the first term is Fisher information matrix. As a result, the Fisher information matrix is the Hessian matrix of the Kullback--Leibler distance evaluated at the true parameter $\bm\theta$.
\begin{equation}
\nabla^2_{\bm\theta'}\mathbf{D}({\bm\theta}||{\bm\theta'})\vert_{\bm\theta'=\bm\theta} = \mathbf{I}(\bm\theta).
\label{eq:kl8}
\end{equation}


\bibliographystyle{model3-num-names}
\bibliography{allrefs}

\end{document}